\documentclass[journal,twoside]{IEEEtran}
\usepackage[margin=2.2cm, centering]{geometry}
\usepackage{multirow}
\usepackage{booktabs}
\usepackage{lipsum}
\usepackage{amsfonts}
\usepackage{placeins}
\usepackage{array}
\usepackage{amsmath,cases,amssymb}
\usepackage{graphicx}
\usepackage{cite}
\usepackage{subfigure}
\usepackage{epsfig}
\usepackage{url}
\usepackage{algorithm}
\usepackage{algorithmic}
\usepackage{epstopdf}
\usepackage{balance}
\usepackage{color}
\newcommand{\lbn}{\par\medskip\noindent}

\newcommand{\ds}{\displaystyle}
\newtheorem{myth}{Theorem}

\newcommand{\la}{\langle}
\newcommand{\ra}{\rangle}

\newcommand{\st}{{\mathrm{subject\ to}}}
\newcommand{\maxi}{{\mathrm{maximize}}}

\newcommand{\clL}{{\cal L}}

\newcommand{\clJ}{{\cal J}}

\newcommand{\clA}{{\cal A}}
\newcommand{\clR}{{\cal R}}
\newcommand{\clM}{{\mathbf{\mathcal{ M}}}}

\newcommand{\clQ}{{\cal Q}}

\newcommand{\bH}{\mathbf{H}}
\newcommand{\bV}{\mathbf{V}}

\begin{document}

\title{Precoder Design for Signal Superposition in MIMO-NOMA Multicell Networks
\thanks{Part of this work has been submitted to IEEE Global Communication Conference (GLOBECOM 2017), Singapore, Dec. 2017.}}
\author{Van-Dinh Nguyen, \textit{Student Member, IEEE,} Hoang Duong Tuan,  Trung Q. Duong, \textit{Senior Member, IEEE,} \\ H. Vincent Poor, \textit{Fellow, IEEE,}   and Oh-Soon Shin, \textit{Member, IEEE}
 \thanks{V.-D. Nguyen and O.-S. Shin are with the School of Electronic Engineering, Soongsil University, Seoul 06978, Korea (e-mail: \{nguyenvandinh, osshin\}@ssu.ac.kr).}
\thanks{H. D. Tuan is with the Faculty of Engineering and Information Technology,
University of Technology, Sydney, NSW 2007, Australia (email:
tuan.hoang@uts.edu.au).}
\thanks{T. Q. Duong is with the School of Electronics, Electrical Engineering and Computer Science, Queen's University Belfast, Belfast BT7 1NN, United Kingdom (e-mail: trung.q.duong@qub.ac.uk).}
\thanks{H. V. Poor is with Department of Electrical Engineering, Princeton University, Princeton, NJ 08544 USA (e-mail: poor@princeton.edu).}}
\markboth{IEEE JOURNAL ON SELECTED AREAS IN COMMUNICATIONS, ACCEPTED FOR PUBLICATION}%
{IEEE JOURNAL ON SELECTED AREAS IN COMMUNICATIONS, ACCEPTED FOR PUBLICATION} 
\maketitle
\begin{abstract}
The throughput of users with poor channel conditions, such as those at a cell edge, 
 is a bottleneck  in wireless  systems. A major part  of the  power budget must be allocated to serve these users
in guaranteeing  their quality-of-service (QoS) requirement, hampering QoS for other users and thus compromising
the system reliability. In nonorthogonal multiple access (NOMA), the message intended for a user with a
poor channel condition is decoded by itself and by another user with a better channel condition.
The message intended for the latter is then successively decoded by itself after canceling the interference of the former.
The overall information throughput is thus improved by this
particular successive  decoding and interference cancellation.
This paper aims to design  linear precoders/beamformers for signal superposition at the base stations of NOMA  multi-input multi-output  multi-cellular systems to maximize  the overall sum throughput subject to
the users' QoS requirements, which are imposed independently on
the users' channel condition. This design problem is formulated as the maximization of a highly nonlinear and nonsmooth function subject to nonconvex constraints,
which is very computationally challenging. Path-following algorithms for its solution, which
invoke only a simple convex problem of moderate dimension at each iteration are developed.
Generating a sequence of improved points, these algorithms  converge at least to a local optimum.
Extensive numerical simulations are then provided to demonstrate their merit.
\end{abstract}
\begin{IEEEkeywords}
Multi-user interference system, multi-input multi-output (MIMO),   nonorthogonal multiple access (NOMA), nonconvex optimization,  quality-of-service (QoS), successive interference cancellation (SIC), signal superposition.
\end{IEEEkeywords}

\section{Introduction}\label{sec:intro}
The explosive  growth of mobile traffic demand  is the driving force
behind the development of signal processing and communication technologies to
significantly upgrade the  high-end experiences of
communication   such as high throughput, high reliability, and ubiquitous access.
 It is widely believed that interference-limited techniques such as
coordinated multipoint  transmission (CoMP) \cite{GH10,JYH12}, which treat interference as noise
cannot meet the edge throughput requirements for 5G \cite{Anetal14}. Dirty-paper coding (DPC) \cite{WeingartenIT06}, under which the interference is successively mitigated, can improve both the edge and sum throughput  but is difficult to implement in practice and  remains only as a theoretical concept  due to its high computational complexity. Nonorthogonal multiple access (NOMA)
has been recently recognized as an essential enabling technology for 5G systems \cite{DingMag16} due to its potential
to improve the edge throughput \cite{Saietal13}.

In NOMA, a base station (BS) transmits a signal superposition to all users. The users are paired so that
in each pair there is one with a better channel condition and another with a poorer channel condition \cite{Benjebbour14}.
The messages intended for each pair of users are sequentially decoded as follows. First,
the message for the user with the poorer channel condition is  decoded by
both users. The message for the user with the better channel condition is then successively decoded
by this user after canceling  the interference from the other user  \cite{DingTVT15}. Thus, while the  throughput
at the users with poorer channel condition remains  the same as that in interference-limited techniques, the throughput
at the users with better channel condition is  clearly improved, leading to a higher system throughput.

Multi-input multi-output (MIMO) is widely known for its enormous potential in improving the capacity of wireless communication systems without requiring extra bandwidth or  power.  NOMA for MIMO communication
(MIMO-NOMA) in single-cell systems for achieving higher throughput has been investigated in \cite{DingTWC16,DSP16}, and an extension  to multi-cell cases has been considered in \cite{ShinComL16}. In  multi-cell systems, the effects of inter-cell interference are acute and unpredictable,  limiting the quality-of-service (QoS) for cell-edge users. It is therefore challenging to realize the benefit that NOMA may bring to  multi-cell systems.

\subsection{Related Works}
In this subsection we discuss  the state-of-the-art of signal processing techniques
for NOMA downlink transmission. NOMA was mostly studied for single-cell
multi-input single-output (MISO) systems known as MISO-NOMA, where the multiple-antenna BS
broadcasts signal superpositions to single-antenna users.  Under the assumption on low QoS requirement for
the near user (with a good channel condition) and high QoS requirement for the far user (with a poor channel condition) in
a two-user MISO-NOMA, \cite{Choi15} proposed a  heuristic computational procedure with neither convergence nor
optimality guaranteed  for the beamforming power minimization. Under
similar users' QoS requirements in a $2K$-user MISO-NOMA, it used a particular zero-forcing beamformer
to cancel the inter-pair interference, so   the problem of $2K$-user MISO-NOMA beamforming is decomposed
into $K$ independent subproblems of two-user MISO-NOMA beamforming. A closed-form solution
for minimization of beamforming power in  two-user MISO-NOMA subject to  natural users' QoS requirements
was obtained in \cite{ChenCL16,ChenTSP16,ChenAC16}. In  \cite{HDRK16}, users  performed
successive interference cancellation (SIC) based on the channel gain differences.
Its proposed algorithm for beamforming is of high computational complexity.

Regarding  MIMO-NOMA, \cite{DingTWC16} and \cite{DSP16} derived  the outage probability experienced
by users in zero-forcing postcoding or signal alignment. Power allocation for achieving the ergodic capacity of
two-user MIMO NOMA was considered in  \cite{SunWCL15} and \cite{ChoiTWC16}.
User-pairing to enhance the throughput of users of poor channel condition was proposed in \cite{LuiCL16}.
 \cite{ShinComL16}  proposed two interference alignment based coordinated beamforming
for a two-cell MIMO-NOMA, where the interference at all cell-center users and edge-center users is canceled.

\subsection{Motivation and Contributions }
The paper considers the problem of  designing linear precoders/beamformers at the BSs for MIMO-NOMA multi-cell systems
to maximize their sum throughput while simultaneously meeting the users' QoS requirements.
In general, such a design problem is very complicated as
the objective function is  nonlinear and nonsmooth, and the QoS constraints are highly nonconvex, for which
even  finding a feasible point is already challenging.
The main contributions of the paper are three-fold:

\begin{itemize}
   \item For MIMO-NOMA,  two path-following optimization algorithms are proposed for computation, which
   at least converge to a locally optimal solution.
    At each iteration, the first algorithm invokes a convex quadratic program  while the second
    algorithm invokes a semi-definite program (SDP). Both these convex programs are of moderate dimension so
    their computation is very efficient.
	
	\item Another path-following algorithm tailored for MISO-NOMA is proposed, which explores much simpler structures of
the throughput functions in MISO systems for more efficient computation.
		
	\item The provided numerical results show the essential performance improvements of NOMA based systems compared to
the conventional systems. The capability of NOMA to improve both edge and sum throughput is revealed.
	\end{itemize}

\subsection{ Paper Organization and Notation}

The rest of the paper is organized as follows.  Section~\ref{sec:sys_model} presents the system model and formulates the problem. A convex quadratic programming based path-following algorithm for the MIMO-NOMA  problem is developed in Section~\ref{sec:CQBI}, while another
SDP based path-following algorithm  is developed in Section~\ref{SDPBI}. Section~\ref{sec:TL} devotes to computation for MISO-NOMA problem. Numerical results are provided in Section~\ref{sec:simulation}, and Section~\ref{sec:conclusion} concludes the paper.

\textit{Notation.} Bold-faced upper-case letters are used for matrices, bold-faced lower-case letters
 are used for vectors, and lower-case letters are  used for for scalars.
$\mathbf{I}_n$ is the identity matrix of size $n\times n$.
 $\mathbf{X}^{H}$, $\mathbf{X}^{T}$, and $\mathbf{X}^{*}$  are the Hermitian transpose, normal transpose,  and conjugate of a matrix $\mathbf{X}$, respectively. The inner product $\langle{\mathbf{X}, \mathbf{Y}}\rangle$ of matrices $\mathbf{X}$ and $\mathbf{Y}$
 is defined as $\textrm{trace}(\mathbf{X}^H \mathbf{Y})$. Denote by $\langle \mathbf{A}\rangle$  the trace of a matrix $\mathbf{A}$, and by $|\mathbf{A}|$ its determinant. $\|\cdot\|$  stands for matrix's Frobenius norm
 or vector's Euclidean norm.
 For Hermitian symmetric matrices $\mathbf{A}\succeq \mathbf{0}$ ($\mathbf{A}\succ \mathbf{0}$, resp.) means that
 $\mathbf{A}$ is a positive semidefinite (positive definite, resp.) matrix. Accordingly
$\mathbf{A} \succeq \mathbf{B}$ ($\mathbf{A} \succ \mathbf{B}$, resp.) means
 $\mathbf{A}-\mathbf{B}\succeq \mathbf{0}$   ($\mathbf{A}-\mathbf{B}\succ \mathbf{0}$, resp.).
 $\mathbb{C}$ is the set of all complex numbers, and $\emptyset$ is an empty set. $\Re\{x\}$ denotes the real part of a complex number $x$. $\mathbf{x}\sim\mathcal{CN}(\boldsymbol{\eta},\boldsymbol{Z})$ means that $\mathbf{x}$ is a random vector following a complex circular Gaussian distribution with mean  $\boldsymbol{\eta}$ and covariance matrix $\boldsymbol{Z}$. $\nabla_{\mathbf{x}}f(\mathbf{x})$ is the gradient of function  $f(\cdot)$ with respect to its variable $\mathbf{x}$. $\mathbb{E}\{\cdot\}$ denotes the expectation operator.

\section{System Model and Problem Formulation}\label{sec:sys_model}

This section presents a system model for signal superposition in MIMO-NOMA multi-cell systems
and formulates an optimization problem for the precoder design. Relations
to CoMP and DPC  are also briefly clarified.

\subsection{Signal Processing Model}

\begin{figure}[t]
\centering
\includegraphics[width=0.6\textwidth,trim={0.5cm 0.0cm -4.5cm -0.0cm}]{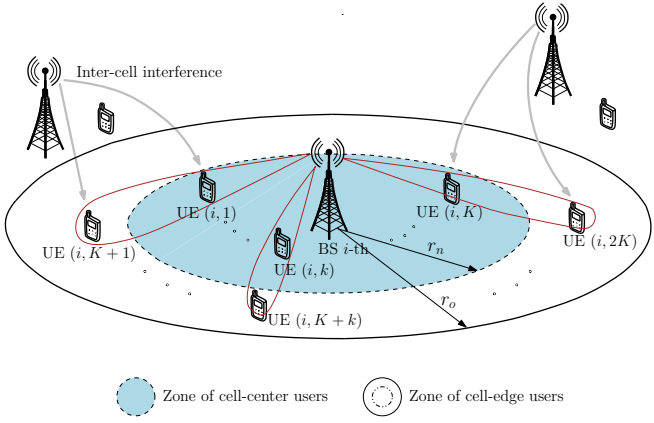}
\caption{An illustration of the cell of interest  in a NOMA system serving $2K$ users.}
\label{fig:SM:1}
\end{figure}

Consider a downlink system consisting of $N$ cells, where
the BS of each cell is equipped with $N_t$ antennas to serve $2K$ users
(UEs) within its cell as illustrated in Fig.~\ref{fig:SM:1}. Each UE is equipped with $N_r$ antennas.  In each cell,  there are $K$ near UEs
(cell-center UEs), which are located inside the circular area with radius $r_n$ and the BS at its center,
and $K$ far UEs (cell-edge UEs), which are located within the ring area with inner radius $r_n$ and
outer radius $r_o$. $K$ far UEs in each cell are not only in poorer
channel conditions than other $K$ near UEs but also are under
more intensified  inter-cell interference from adjacent cells.

Upon denoting $\mathcal{I} \triangleq \{1, 2, \cdots, N\}$ and $\mathcal{J}  \triangleq  \{1, 2, \cdots, 2K\}$,
the $j$-th UE in the $i$-th cell is referred to as UE $(i,j)\in \mathcal{S}\triangleq \mathcal{I}\times\mathcal{J}$.
The cell-center UEs are UE $(i,j)$, $j\in\mathcal{K}_1\triangleq\{1,\cdots, K\}$ while the cell-edge UEs are
UE $(i,j)$, $j\in\mathcal{K}_2\triangleq\{K+1,\cdots, 2K\}$. Thus the set of cell-center UEs and the set of cell-edge UEs
are $\mathcal{S}_1\triangleq \mathcal{I}\times \mathcal{K}_1$ and $\mathcal{S}_2\triangleq \mathcal{I}\times \mathcal{K}_2$,
respectively. In NOMA each
cell-center UE $(i,j)\in\mathcal{S}_1$ is randomly paired with cell-edge UE $(i,p(j))\in \mathcal{S}_2$ of the same cell to create
a virtual cluster.\footnote{Using more sophisticated user-pairing strategies may improve the performance of   MIMO-NOMA networks (see e.g. \cite{DingTVT15,LuiCL16}) but it is beyond the scope of this paper.} The signal superpositions  are precoded at the BSs prior to being transmitted to the UEs. Specifically, the message intended for UE $(i,j)$ is
$\mathbf{s}_{i,j} \in \mathbb{C}^L$ with $\mathbb{E}\{\mathbf{s}_{i,j}(\mathbf{s}_{i,j})^H\}=\mathbf{I}_L$, which
is precoded by
matrix $\mathbf{V}_{i,j}\in \mathbb{C}^{N_t\times L}$, where $L$ is the number of concurrent data streams and $L \leq \min\{N_t,N_r\}$.
For notational convenience, let us define
$\mathbf{V} \triangleq  [\mathbf{V}_{i,j}]_{(i,j)\in\mathcal{S}}$. The received signals at UE $(i,j)$ and UE $(i,p(j))$ are expressed as
\begin{IEEEeqnarray}{rCl}
    \mathbf{y}_{i,j} &=& \mathbf{H}_{i,i,j}\mathbf{V}_{i,j}\mathbf{s}_{i,j}+\mathbf{H}_{i,i,j}\mathbf{V}_{i,p(j)}\mathbf{s}_{i,p(j)} \nonumber\\
   & & + \ds\sum_{(s,l)\in {\cal S}\setminus \{(i,j), (i,p(j)) \}} \mathbf{H}_{s,i,j}\mathbf{V}_{s,l} \mathbf{s}_{s,l} + \mathbf{n}_{i,j},\label{eq:yij}
 \end{IEEEeqnarray}
and
\begin{IEEEeqnarray}{rCl}
        && \mathbf{y}_{i,p(j)} = \mathbf{H}_{i,i,p(j)}\mathbf{V}_{i,p(j)}\mathbf{s}_{i,p(j)}+\mathbf{H}_{i,i,p(j)}\mathbf{V}_{i,j}\mathbf{s}_{i,j}\nonumber\\
				&&\qquad +    \ds\sum_{(s,l)\in {\cal S}\setminus \{(i,j), (i,p(j)) \}} \mathbf{H}_{s,i,p(j)}\mathbf{V}_{s,l} \mathbf{s}_{s,l} + \mathbf{n}_{i,p(j)}\quad\label{eq:yipj}
\end{IEEEeqnarray}
where $\mathbf{H}_{s,i,j} \in\mathbb{C}^{N_r \times N_t}$ is the MIMO channel from the BS $s\in\mathcal{I}$ to UE $(i,j)\in\mathcal{S}$. The entries of the additive noise  $\mathbf{n}_{i,j}\in\mathbb{C}^{N_r}$ are independent and identically
distributed (i.i.d.) noise samples with zero mean and variance $\sigma^2$.
The covariances of $\mathbf{y}_{i,j}$ and $\mathbf{y}_{i,p(j)}$ are thus
\begin{equation}\label{eqLMij}
\clM_{i,j}(\mathbf{V})= \ds\sum_{(s,l)\in{\cal S}}\mathbf{H}_{s,i,j}\mathbf{V}_{s,l}\mathbf{V}_{s,l}^H \mathbf{H}_{s,i,j}^H + \sigma^2\mathbf{I}_{N_r},
\end{equation}
and
\begin{equation}\label{eqLMipj}
\clM_{i,p(j)}(\mathbf{V})= \ds\sum_{(s,l)\in{\cal S}}\mathbf{H}_{s,i,p(j)}\mathbf{V}_{s,l}\mathbf{V}_{s,l}^H \mathbf{H}_{s,i,p(j)}^H + \sigma^2\mathbf{I}_{N_r}.
\end{equation}
The purpose of the paper is to design complex-valued precoding matrices $\bV_{i,j}$ to maximize the overall
spectral efficiency under a given pairing for NOMA. The MIMO channel states $\mathbf{H}_{s,i,j}$
are assumed unchanged during message transmission but may change independently from one message to another
and are perfectly known at all nodes \cite{Choi15,HDRK16,DSP16}.

\subsection{Problem Formulation}
In NOMA, the message $\mathbf{s}_{i,p(j)}$ intended for the cell-edge UE $(i,p(j))$ is decoded by  both UE $(i, p(j))$ and
UE $(i,j)$. Then  message $\mathbf{s}_{i,j}$ intended for the cell-center   UE $(i,j)$ is decoded by  itself only.

The cell-edge UE $(i,p(j))$ decodes its own message $\mathbf{s}_{i,p(j)}$ with the achievable rate
\begin{IEEEeqnarray}{rCl}\label{eq:Rijj}
         &&    \clR_{i,p(j)}^{p(j)}(\mathbf{V}) = \ln\Bigl| \mathbf{I}_L +  \nonumber\\
				 &&\quad		  (\mathbf{V}_{i,p(j)})^H\mathbf{H}_{i,i,p(j)}^H
            \clM_{i,p(j)}^{p(j)}(\mathbf{V})^{-1}\mathbf{H}_{i,i,p(j)}\mathbf{V}_{i, p(j)} \Bigl| \qquad
\end{IEEEeqnarray}
where $\clM_{i,p(j)}^{p(j)}(\mathbf{V})$ is defined by
\begin{IEEEeqnarray}{rCl}\label{eq:Mij}
&&\clM_{i,p(j)}^{p(j)}(\mathbf{V})  \nonumber\\
&&\ \triangleq \clM_{i,p(j)}(\mathbf{V}) - \mathbf{H}_{i,i,p(j)}\mathbf{V}_{i,p(j)}(\mathbf{V}_{i,p(j)})^H\mathbf{H}_{i,i,p(j)}^H \nonumber\\
&&\ =\ds\sum_{(s,l)\in{\cal S}\setminus\{(i,p(j)) \}}\mathbf{H}_{s,i,p(j)}\mathbf{V}_{s,l}\mathbf{V}_{s,l}^H \mathbf{H}_{s,i,p(j)}^H + \sigma^2\mathbf{I}_{N_r}.\quad\,
\end{IEEEeqnarray}
On the other hand, the cell-center UE $(i,j)$ decodes the message $\mathbf{s}_{i,p(j)}$ with the achievable rate
\begin{IEEEeqnarray}{rCl}\label{eq:Rijpj}
&&\clR_{i,p(j)}^j(\mathbf{V}) = \nonumber\\
&&\ \; \ln\Bigl|\mathbf{I}_L + (\mathbf{V}_{i,p(j)})^H\mathbf{H}_{i,i,j}^H\clM_{i,j}^{p(j)}(\mathbf{V})^{-1}\mathbf{H}_{i,i,j}\mathbf{V}_{i,p(j)}\Bigl|\quad
 \end{IEEEeqnarray}
where
\begin{IEEEeqnarray}{rCl}\label{eq:Mijpj}
&&\clM_{i,j}^{p(j)}(\mathbf{V})  \triangleq  \clM_{i,j}(\mathbf{V}) - \mathbf{H}_{i,i,j}\mathbf{V}_{i,p(j)}(\mathbf{V}_{i,p(j)})^H\mathbf{H}_{i,i,j}^H\quad\nonumber\\
&&\qquad=\ds\ds\sum_{(s,l)\in{\cal S}\setminus\{(i,p(j))\}}\mathbf{H}_{s,i,j}\mathbf{V}_{s,l}\mathbf{V}_{s,l}^H \mathbf{H}_{s,i,j}^H + \sigma^2\mathbf{I}_{N_r}.
\end{IEEEeqnarray}
Hence, the throughput by decoding the message $\mathbf{s}_{i,p(j)}$ by UEs $(i,p(j))$ and $(i,j)$ is
\begin{equation}\label{noma1}
\clR_{i,p(j)}=\min \Bigl\{\clR^{j}_{i,p(j)}(\mathbf{V}), \clR^{p(j)}_{i,p(j)}(\mathbf{V})\Bigr\}.
\end{equation}
Next, the  throughput by decoding the message $\mathbf{s}_{i,j}$
by the cell-center UE $(i,j)$ after decoding the message $\mathbf{s}_{i,p(j)}$ is
\begin{equation}\label{noma2}
            \clR_{i,j}(\mathbf{V}) = \ln\Bigl| \mathbf{I}_L + (\mathbf{V}_{i,j})^H\mathbf{H}_{i,i,j}^H\clM_{i,j}^{p}(\mathbf{V})^{-1}\mathbf{H}_{i,i,j}\mathbf{V}_{i,j} \Bigl|
 \end{equation}
where
\begin{IEEEeqnarray}{rCl}\label{eq:Mipjpj}
       && \clM_{i,j}^{p}(\mathbf{V})  \triangleq  \clM_{i,j}^{p(j)}(\mathbf{V}) - \mathbf{H}_{i,i,j}\mathbf{V}_{i,j}(\mathbf{V}_{i,j})^H\mathbf{H}_{i,i,j}^H \nonumber\qquad\quad\\
       && =\ds\ds\sum_{(s,l)\in{\cal S}\setminus\{(i,p(j)),(i,j)\}}\mathbf{H}_{s,i,j}\mathbf{V}_{s,l}\mathbf{V}_{s,l}^H \mathbf{H}_{s,i,j}^H + \sigma^2\mathbf{I}_{N_r}.
 \end{IEEEeqnarray}

Our  goal is to maximize the total sum throughput  of the system under QoS for each individual UE and
 power budget at each BS, which is mathematically formulated as
\begin{IEEEeqnarray}{rCl}\label{Rmtgor}
        &&\underset{\mathbf{V}}{\maxi}\; \mathcal{P}(\mathbf{V})\triangleq\sum_{i=1}^N\sum_{j=1}^K\Bigl(\clR_{i,j}(\mathbf{V})+\clR_{i,p(j)}(\mathbf{V})\Bigr)\qquad \IEEEyessubnumber\label{Rmtgora}\\
        &&\st\;  \clR_{i,j}(\mathbf{V}) \geq r_{i,j},\  \forall i \in\mathcal{I},\ \forall j\in\mathcal{K}_1, \IEEEyessubnumber\label{Rmtgorb}\\
        &&\qquad\qquad\;  \clR_{i,p(j)}(\mathbf{V}) \geq r_{i,p(j)},\,  \forall i \in\mathcal{I},\, \forall j\in\mathcal{K}_2, \IEEEyessubnumber\label{Rmtgorc}\\
        &&\qquad\qquad\;    \sum_{j\in \clJ} \Bigl\langle \mathbf{V}_{i,j}\mathbf{V}_{i,j}^H  \Bigl\rangle \le P^{\max}_{i},\ \forall i \in\mathcal{I}\IEEEyessubnumber\label{Rmtgord}
 \end{IEEEeqnarray}
where $P^{\max}_i$ in \eqref{Rmtgord} is the transmit power budget of BS $i$. The QoS constraints
\eqref{Rmtgorc} and \eqref{Rmtgord} set a minimum throughput requirement $r_{i,j}$ at the UE $(i,j)$ and  $r_{i,p(j)}$ at the UE $(i,p(j))$.

\subsection{Relations to  CoMP and DPC}
In CoMP \cite{GH10,JYH12},  the problem of maximizing the sum throughput under QoS constraints is formulated as \cite{TTN16}
\begin{IEEEeqnarray}{rCl}\label{CoMPobj}
        &&\underset{\mathbf{V}}{\maxi}\;\ \mathcal{P}^{\mathrm{CoMP}}(\mathbf{V})\triangleq\sum_{i=1}^N\sum_{j=1}^{2K}\clR_{i,j}^{'}(\mathbf{V})\IEEEyessubnumber\label{CoMPa}\\
        &&\st\;\  \clR_{i,j}^{'}(\mathbf{V}) \geq r_{i,j},\  \forall i \in\mathcal{I},\ \forall j\in\mathcal{J}, \IEEEyessubnumber\label{CoMPb}\qquad\\
        &&\qquad\qquad\quad    \sum_{j\in \clJ} \Bigl\langle \mathbf{V}_{i,j}\mathbf{V}_{i,j}^H  \Bigl\rangle \le P^{\max}_{i},\ \forall i \in\mathcal{I}\IEEEyessubnumber\label{CoMPc}
  \end{IEEEeqnarray}
where $\clR_{i,j}^{'}(\mathbf{V})$ is given by
\begin{equation}\label{eq:Rij:CoMP}
\clR_{i,j}^{'}(\mathbf{V}) =  \ln\Bigl|\mathbf{I}_L + (\mathbf{V}_{i,j})^H\mathbf{H}_{i,i,j}^H\clM_{i,j}^{j}(\mathbf{V})^{-1}\mathbf{H}_{i,i,j}\mathbf{V}_{i,j}\Bigl|
 \end{equation}
with
\begin{equation}
\begin{array}{lll}
\clM_{i,j}^{j}(\mathbf{V}) \triangleq \ds\ds\sum_{(s,l)\in{\cal S}\setminus\{(i,j)\}}\mathbf{H}_{s,i,j}\mathbf{V}_{s,l}\mathbf{V}_{s,l}^H \mathbf{H}_{s,i,j}^H + \sigma^2\mathbf{I}_{N_r}.
\end{array}
\end{equation}
Compared $\clR_{i,j}^{'}(\mathbf{V})$ defined by $(\ref{eq:Rij:CoMP})$ to
$\clR_{i,p(j)}(\mathbf{V})$ and $\clR_{i,j}(\mathbf{V})$ defined by  (\ref{noma1}) and (\ref{noma2}) one can see that
\begin{IEEEeqnarray}{rCl}
\clR_{i,p(j)}^{'}(\mathbf{V})&=&\clR^{p(j)}_{i,p(j)}(\mathbf{V})\label{com1}\\
 &\geq&\min \Bigl\{\clR^{j}_{i,p(j)}(\mathbf{V}), \clR^{p(j)}_{i,p(j)}(\mathbf{V})\Bigr\}\label{com2}\\
 &=& \clR_{i,p(j)},\ \forall j\in\mathcal{K}_2,\label{com3}
\end{IEEEeqnarray}
and
\begin{equation}
\clR_{i,j}^{'}(\mathbf{V})\leq \clR_{i,j}(\mathbf{V}),\ \forall j\in\mathcal{K}_1,\label{com4}
\end{equation}
i.e., under the same precoder $\mathbf{V}$, the  throughput at cell-edge UEs is higher with CoMP while
that at  cell-center UEs is higher with NOMA. Thus, under the same precoder, NOMA does not need to perform better
than CoMP in terms of the total  throughput. The preference of  NOMA now critically depends on its performance at its
optimal precoder, which is sought in the next sections.

On the other hand, DPC at the BS $i$ with encoding order from UE $(i,2K)$ to UE $(i,1)$
enables  UE $(i,j)$ view the massages intended for   UEs $(i,j')$, $j'>j$ as non-causually known  and
thus cancel them from its received signal.
Hence, the  throughput by decoding the message $\mathbf{s}_{i,j}$ for  UE $(i,j)$  is defined by
\begin{equation}\label{DPC1}
            \clR_{i,j}^{''}(\mathbf{V}) = \ln\Bigl|\mathbf{I}_L  + (\mathbf{V}_{i,j})^H\mathbf{H}_{i,i,j}^H\clM_{i,j}^{i}(\mathbf{V})^{-1}\mathbf{H}_{i,i,j}\mathbf{V}_{i,j} \Bigl|
 \end{equation}
where
\begin{IEEEeqnarray}{rCl}\label{eq:DPC2}
        \clM_{i,j}^{i}(\mathbf{V})  &\triangleq&    \ds\ds\sum_{s \neq i}^{N}\sum_{l = 1}^{2K}\mathbf{H}_{s,i,j}\mathbf{V}_{s,l}\mathbf{V}_{s,l}^H \mathbf{H}_{s,i,j}^H   \nonumber\\
			&+&\;	\sum_{k=1}^{j-1}\mathbf{H}_{i,i,j}\mathbf{V}_{i,k}\mathbf{V}_{i,k}^H\mathbf{H}_{i,i,j}^H + \sigma^2\mathbf{I}_{N_r}.\quad
 \end{IEEEeqnarray}
The problem of maximizing the sum throughput under QoS constraints is formulated as
\begin{IEEEeqnarray}{rCl}\label{DPCobj}
        &&\underset{\mathbf{V}}{\maxi}\;\ \mathcal{P}^{\mathrm{DPC}}(\mathbf{V})\triangleq\sum_{i=1}^N\sum_{j=1}^{2K}\clR_{i,j}^{''}(\mathbf{V})\IEEEyessubnumber\label{DPCa}\\
        &&\st\;\  \clR_{i,j}^{''}(\mathbf{V}) \geq r_{i,j},\  \forall i \in\mathcal{I},\ \forall j\in\mathcal{J}, \IEEEyessubnumber\label{DPCb}\qquad\\
        &&\qquad\qquad\quad    \sum_{j\in \clJ} \Bigl\langle \mathbf{V}_{i,j}\mathbf{V}_{i,j}^H  \Bigl\rangle \le P^{\max}_{i},\ \forall i \in\mathcal{I}.\IEEEyessubnumber\label{DPCc}
  \end{IEEEeqnarray}
Apparently, under DPC the intra-cell interference from UEs of better channel
condition is canceled in decoding the messages intended for UEs of poorer channel condition resulting in much better
edge throughput compared with CoMP. NOMA is advantageous over DPC in terms of ease of implementation.
By increasing the QoSs' requirement in
(\ref{Rmtgorb}), (\ref{CoMPb}) and (\ref{DPCb}), we will see that the NOMA's sum throughput is fully superior over CoMP's one and catches  up DPC's one.
\section{Convex Quadratic Based Iterations}\label{sec:CQBI}
Note that the QoS constraints \eqref{Rmtgorb} and \eqref{Rmtgorc} in \eqref{Rmtgor},
and \eqref{CoMPb} in \eqref{CoMPobj} are set beforehand, which are dependent on the UEs' throughput requirements but
independent on their channel condition. In maximizing the  sum throughput objective \eqref{CoMPa}, most of CoMP techniques
(see e.g. \cite{CACC08}) are unable to address the QoS constraints \eqref{CoMPb}. Without setting such QoS constraints,
the UEs of poor channel condition are easily disconnected from the service because
it is well known that almost all of transmit power will be
allocated to a very few UEs of the best channel conditions in maximizing the sum throughput, causing almost zero throughput at other UEs. The weighted sum throughput maximization is  only an ad hoc way to
balance  the UEs' throughput. Alternatively, the throughput satisfaction can be effectively handled via maximizing the
users' worst throughput but the latter involves optimization of a nonsmooth objective function, for which these techniques are powerless. Both QoS constrained  sum throughput maximization problem \eqref{CoMPobj} and UEs' worst throughput maximization
problem could be addressed very recently in \cite{TTN16}.

Although the optimization problem \eqref{Rmtgor} is different from \eqref{CoMPobj} and \eqref{DPCobj}, all functions appearing in the former
have a similar structure to that appearing in the latter. Therefore, a systematic approach to solve the latter is expected
to be applicable to the former.  In this section, we adopt the approach of \cite{TTN16} to address \eqref{Rmtgor}. Unlike
\cite{Cetal15}, which aims at expressing the nonsmooth function $\clR_{i,p(j)}(\mathbf{V})$ in \eqref{noma1} and then
the nonsmooth objective function in \eqref{Rmtgorb} as d.c. (\underline{d}ifference of two \underline{c}onvex functions)
by using the universality of d.c. functions \cite{Tuybook} and leads to d.c. iterations \cite{KTN12}  of high computational complexity, we will see now that each below iteration invokes  only a simple convex quadratic
program of low computational complexity.

Suppose that $\mathbf{V}^{(\kappa)}\triangleq [\mathbf{V}^{(\kappa)}_{i,j}]_{(i,j)\in\mathcal{S}}$ is a feasible
point found at the $(\kappa-1)$-th iteration. Define the following
quadratic functions in $\bV$:
\begin{IEEEeqnarray}{rCl}
&&\clR^{j,(\kappa)}_{i,p(j)}(\mathbf{V})
\triangleq a_{i,p(j)}^{j,(\kappa)}+2\Re\Bigl\{\bigl\la\clA_{i,p(j)}^{j,(\kappa)},\mathbf{V}_{i,p(j)}\bigl\ra
\Bigl\}\nonumber\\
&&\quad -\,\Bigl\la\clM_{i,j}^{p(j)}(\mathbf{V}^{(\kappa)})^{-1}-\clM_{i,j}(\mathbf{V}^{(\kappa)})^{-1},\clM_{i,j}(\mathbf{V})\Bigl\ra,
\nonumber\\
&&\clR^{p(j),(\kappa)}_{i,p(j)}(\mathbf{V}) \triangleq
a_{i,p(j)}^{p(j),(\kappa)}
+2\Re\Bigl\{\bigl\la \clA_{i, p(j)}^{p(j),(\kappa)},\mathbf{V}_{i, p(j)}\bigl\ra
\Bigl\}\nonumber\\
&&\quad-\,\Bigl\la \clM_{i,p(j)}^{p(j)}(\mathbf{V}^{(\kappa)})^{-1}-\clM_{i,p(j)}(\mathbf{V}^{(\kappa)})^{-1},
\clM_{i,p(j)}(\mathbf{V})\Bigl\ra,\label{eq:qdt}  \nonumber     \\
&&\clR^{(\kappa)}_{i,j}(\mathbf{V}) \triangleq a_{i,j}^{(\kappa)}
+2\Re\Bigl\{\bigl\la\clA_{i,j}^{(\kappa)},\mathbf{V}_{i,j}\bigl\ra\Bigl\}\nonumber \\
&&\quad-\,\Bigl\la \clM_{i,j}^{p}(\mathbf{V}^{(\kappa)})^{-1}-\clM_{i,j}^{p(j)}(\mathbf{V}^{(\kappa)})^{-1},
\clM_{i,j}^{p(j)}(\mathbf{V})\Bigl\ra
\end{IEEEeqnarray}
where $a_{i,p(j)}^{j,(\kappa)}$, $a_{i,p(j)}^{p(j),(\kappa)}$, $a_{i,j}^{(\kappa)}$ are given as
\begin{IEEEeqnarray}{rCl}
 a_{i,p(j)}^{j,(\kappa)} &\triangleq& \clR^{j}_{i,p(j)}(\mathbf{V}^{(\kappa)}) - \Bigl\la (\mathbf{V}_{i,p(j)}^{(\kappa)})^H \mathbf{H}_{i,i,j}^H\nonumber\\
    &&\quad \times\clM_{i,j}(\mathbf{V}^{(\kappa)})^{-1}\mathbf{H}_{i,i,j}\mathbf{V}_{i,p(j)}^{(\kappa)}\Bigl\ra\, < 0,\nonumber\\
 a_{i,p(j)}^{p(j),(\kappa)} &\triangleq& \clR^{p(j)}_{i,p(j)}(\mathbf{V}^{(\kappa)}) - \Bigl\la(\mathbf{V}_{i,p(j)}^{(\kappa)})^H\mathbf{H}_{i,i,p(j)}^H \nonumber\\
        &&\quad \times \clM_{i,p(j)}(\mathbf{V}^{(\kappa)})^{-1}\mathbf{H}_{i,i,p(j)}\mathbf{V}_{i,p(j)}^{(\kappa)}\Bigl\ra\, < 0,\nonumber\quad\\
 a_{i,j}^{(\kappa)} &\triangleq& \clR_{i,j}(\mathbf{V}^{(\kappa)}) - \Bigl\la (\mathbf{V}_{i,j}^{(\kappa)})^H\mathbf{H}_{i,i,j}^H  \nonumber \\
 &&\quad \times\clM_{i,j}^{p(j)}(\mathbf{V}^{(\kappa)})^{-1}
\mathbf{H}_{i,i,j}\mathbf{V}_{i,j}^{(\kappa)}\Bigl\ra\, < 0,
\end{IEEEeqnarray}
and $\clA_{i,p(j)}^{j,(\kappa)}$, $\clA_{i,p(j)}^{p(j),(\kappa)}$,  $\clA_{i,j}^{(\kappa)}$ are given as
\begin{IEEEeqnarray}{rCl}
\clA_{i,p(j)}^{j,(\kappa)}&\triangleq& \mathbf{H}_{i,i,j}^H\clM_{i,j}(\mathbf{V}^{(\kappa)})^{-1}\mathbf{H}_{i,i,j}\mathbf{V}_{i,p(j)}^{(\kappa)},\nonumber\\
\clA_{i,p(j)}^{p(j),(\kappa)}&\triangleq&\mathbf{H}_{i,i,p(j)}^H\clM_{i,p(j)}(\mathbf{V}^{(\kappa)})^{-1}
\mathbf{H}_{i,i,p(j)}\mathbf{V}_{i,p(j)}^{(\kappa)},\nonumber  \qquad\\
\clA_{i,j}^{(\kappa)}&\triangleq&\mathbf{H}_{i,i,j}^H\clM_{i,j}^{p(j)}(\mathbf{V}^{(\kappa)})^{-1}
\mathbf{H}_{i,i,j}\mathbf{V}_{i,j}^{(\kappa)}.
\end{IEEEeqnarray}
Note that all functions in (\ref{eq:qdt}) are concave   due
to \eqref{eq:Mij}, \eqref{eq:Mijpj}, and  \eqref{eq:Mipjpj}:
\begin{IEEEeqnarray}{rCl}
\clM_{i,p(j)}^{p(j)}(\mathbf{V}^{(\kappa)})^{-1}-\clM_{i,p(j)}(\mathbf{V}^{(\kappa)})^{-1}&\succeq& \mathbf{0},\nonumber \\
\clM_{i,j}^{p(j)}(\mathbf{V}^{(\kappa)})^{-1}-\clM_{i,j}(\mathbf{V}^{(\kappa)})^{-1}
&\succeq& \mathbf{0},\nonumber \\
\clM_{i,j}^{p}(\mathbf{V}^{(\kappa)})^{-1}-\clM_{i,j}^{p(j)}(\mathbf{V}^{(\kappa)})^{-1} &\succeq& \mathbf{0}.\nonumber
\end{IEEEeqnarray}

The following result shows that the complicated function defined by  \eqref{noma1} and \eqref{noma2}
is lower bounded by concave quadratic functions.

\lbn
\begin{myth}\label{basicth} For $\clR^{(\kappa)}_{i,p(j)}(\mathbf{V}) \triangleq \min\Bigl\{\clR^{j,(\kappa)}_{i,p(j)}(\mathbf{V}),$ $
\clR^{p(j),(\kappa)}_{i,p(j)}(\mathbf{V})\Bigl\}$ it is true that
\begin{IEEEeqnarray}{rCl}\label{main}
\clR_{i,p(j)}(\mathbf{V}^{(\kappa)})&=&\clR^{(\kappa)}_{i,p(j)}(\mathbf{V}^{(\kappa)})\quad\mbox{and}\quad \nonumber \\
\clR_{i,p(j)}(\mathbf{V})&\geq& \clR^{(\kappa)}_{i,p(j)}(\mathbf{V}), \ \forall\, \mathbf{V},
\end{IEEEeqnarray}
and
\begin{IEEEeqnarray}{rCl}\label{mainbound}
\clR_{i,j}(\mathbf{V}^{(\kappa)}) &=& \clR^{(\kappa)}_{i,j}(\mathbf{V}^{(\kappa)})\quad\mbox{and}\quad \nonumber \\
\clR_{i,j}(\mathbf{V}) &\geq& \clR^{(\kappa)}_{i,j}(\mathbf{V}), \ \forall\, \mathbf{V}.
\end{IEEEeqnarray}
\end{myth}
\begin{IEEEproof}
See \cite[Appendix B]{TTN16}.
\end{IEEEproof}\lbn

Based on these results,  at the $\kappa$-th iteration, the following convex program, which is
an inner approximation for the nonconvex optimization problem \eqref{Rmtgor}, is solved to generate the next feasible point
$\mathbf{V}^{(\kappa+1)}$:
          \begin{IEEEeqnarray}{rCl}\label{RmtgFWA}
 &&       \ds\underset{\mathbf{V}}{\max} \quad \mathcal{P}^{(\kappa)}(\mathbf{V}) \triangleq\sum_{i=1}^N\sum_{j=1}^K \Bigl(\clR_{i,j}^{(\kappa)}(\mathbf{V})+\clR_{i,p(j)}^{(\kappa)}(\mathbf{V})  \Bigr)\IEEEyessubnumber\label{RmtgFWA:a}\qquad \\
         &&   \st \ \
 \clR_{i,j}^{(\kappa)}(\mathbf{V})\geq r_{i,j},\ \forall i\in\mathcal{I},\ \forall j\in\mathcal{K}_1,\IEEEyessubnumber\label{RmtgFWA:b}\\
    &&\qquad\qquad\quad    \clR^{(\kappa)}_{i,p(j)}(\mathbf{V})\geq r_{i,p(j)}, \ \forall i\in\mathcal{I},\  \forall j\in\mathcal{K}_2,\IEEEyessubnumber\label{RmtgFWA:c}\\
&&\qquad\qquad\quad    \eqref{Rmtgord}.\IEEEyessubnumber\label{RmtgFWA:d}
        \end{IEEEeqnarray}
A pseudo-code of  this quadratic programing (QP)-based path-following
procedure is given by Algorithm~\ref{alg_SCALE_FW}.

Note that $\mathbf{V}^{(\kappa)}$ is also feasible for \eqref{RmtgFWA} with  $\mathcal{P}(\mathbf{V}^{(\kappa)})=
\mathcal{P}^{(\kappa)}(\mathbf{V}^{(\kappa)})$ by the equalities  in \eqref{main} and \eqref{mainbound}.
It is then true
that $\mathcal{P}^{(\kappa)}(\mathbf{V}^{(\kappa+1)})>\mathcal{P}^{(\kappa)}(\mathbf{V}^{(\kappa)})=\mathcal{P}(\mathbf{V}^{(\kappa)})$ whenever
$\mathbf{V}^{(\kappa+1)}\neq \mathbf{V}^{(\kappa)}$. Together with $\mathcal{P}(\mathbf{V}^{(\kappa+1)})\geq
\mathcal{P}^{(\kappa)}(\mathbf{V}^{(\kappa)})$ and according to the inequalities  in \eqref{main} and \eqref{mainbound}, we have
$\mathcal{P}(\mathbf{V}^{(\kappa+1)})>\mathcal{P}(\mathbf{V}^{(\kappa)})$, i.e., the optimal solution
$\mathbf{V}^{(\kappa+1)}$ of the convex quadratic problem \eqref{RmtgFWA} is a better point for the
nonconvex nonsmooth optimization problem \eqref{Rmtgor} than  $\mathbf{V}^{(\kappa)}$.
Therefore, once initialized from an feasible point $\mathbf{V}^{(0)})$, the sequence $\{\mathbf{V}^{(\kappa)}\}$ obtained by solving \eqref{RmtgFWA} is of improved feasible points for \eqref{Rmtgor}. By following the same arguments as those in \cite[Proposition 1]{TTN16}, we can prove that
 Algorithm~\ref{alg_SCALE_FW}  converges to a Karush-Kuh-Tucker (KKT) point of \eqref{Rmtgor}.
\begin{algorithm}[t]
\begin{algorithmic}[1]
\protect\caption{QP-based path-following algorithm for the STM \eqref{Rmtgor} in MIMO-NOMA}
\label{alg_SCALE_FW}
\global\long\def\algorithmicrequire{\textbf{Initialization:}}
\REQUIRE  Set $\kappa:=0$ and solve \eqref{kiter1}  to generate an initial feasible point $\mathbf{V}^{(0)}$ for constraints
 \eqref{Rmtgorb}-\eqref{Rmtgord}.
\REPEAT
\STATE Solve the convex quadratic program \eqref{RmtgFWA} to obtain the optimal solution: $\mathbf{V}^{\star}$.

\STATE Update\ \ $\mathbf{V}^{(\kappa+1)}:= \mathbf{V}^{\star}.$
\STATE Set $\kappa:=\kappa+1.$
\UNTIL Convergence\\
\end{algorithmic} \end{algorithm}

\textit{Generation of the initial points}:  A feasible point for constraints \eqref{Rmtgorb}-\eqref{Rmtgord}
for initializing  Algorithm~\ref{alg_SCALE_FW} is found via the following problem of QoS feasibility
\begin{IEEEeqnarray}{rCl}\label{qos}
&&\ds\max_{\mathbf{V}} \underset{(i,j)\in\mathcal{S}}{\min}\min \Biggl\{\frac{\clR_{i,j}(\mathbf{V})}{r_{i,j}},
\frac{\clR_{i,p(j)}(\mathbf{V})}{r_{i,p(j)}}\Biggr\}\IEEEyessubnumber\label{qosa}\\
&&\st\quad \eqref{Rmtgord}.\IEEEyessubnumber\label{qosb}
\end{IEEEeqnarray}
Initialized from a feasible point $\mathbf{V}^{(0)}$ for the convex constraint
\eqref{Rmtgord}, the following iterations are invoked
\begin{IEEEeqnarray}{rCl}\label{kiter1}
&&\max_{\mathbf{V}} \underset{(i,j)\in\mathcal{S}}{\min}\min \Biggl\{\frac{\clR^{(\kappa)}_{i,j}(\mathbf{V})}{r_{i,j}},
\frac{\clR^{(\kappa)}_{i,p(j)}(\mathbf{V})}{r_{i,p(j)}}\Biggr\} \IEEEyessubnumber\\
&& \st\quad \eqref{Rmtgord}\IEEEyessubnumber
\end{IEEEeqnarray}
till reaching a value more than or equal to $1$ in satisfying \eqref{Rmtgorb}-\eqref{Rmtgord}.

\textit{Complexity Analysis}:
Problem \eqref{RmtgFWA} is convex quadratic  with
$m_{\text{QP}}=N(1+3K)$ quadratic constraints and $n=KN(2N_tL+1)$ real decision variables. Its
computational complexity is $\mathcal{O}(n^2m_{\text{QP}}^{2.5}+m_{\text{QP}}^{3.5})$.

\section{Semi-Definite Programming Based Iterations}\label{SDPBI}

To further improve the convergence speed of solving \eqref{Rmtgor}, we need explore more partial convex structures of
functions \eqref{noma1} and \eqref{noma2}. In this section, we propose a novel
SDP-based path-following algorithm for \eqref{Rmtgor}. To this end, we will use the following matrix inequalities:
\begin{eqnarray}
\mathbf{V}^H\mathbf{X}^{-1}\mathbf{V} &\succeq& \bar{\boldsymbol{V}}^H\bar{\boldsymbol{X}}^{-1}\mathbf{V}+
\mathbf{V}^H\bar{\boldsymbol{X}}^{-1}\bar{\boldsymbol{V}} \nonumber\\
       && -\; \bar{\boldsymbol{V}}^H\bar{\boldsymbol{X}}^{-1}\mathbf{X}\bar{\boldsymbol{X}}^{-1}\bar{\boldsymbol{V}}, \nonumber\\
			 &&\forall\ \mathbf{V}, \bar{\boldsymbol{V}}, \mathbf{X}\succ  \mathbf{0} , \bar{\boldsymbol{X}}\succ  \mathbf{0} ,  \label{in1}
\end{eqnarray}
and
\begin{eqnarray}
\ln|\mathbf{X}|&\geq& \ln|\bar{\boldsymbol{X}}|-\bigl\la \bar{\boldsymbol{X}}, \mathbf{X}^{-1}-\bar{\boldsymbol{X}}^{-1}\bigl\ra, \nonumber\\
&&\forall\ \mathbf{X}\succ  \mathbf{0} , \bar{\boldsymbol{X}}\succ  \mathbf{0},\label{in2}
\end{eqnarray}
whose proofs are given by Appendix A and Appendix B.

Let us treat the  rate function $\clR_{i,p(j)}^j(\mathbf{V})$ from (\ref{eq:Rijpj}) first.
Applying (\ref{in1}) yields
\begin{IEEEeqnarray}{rCl}
\mathbf{V}_{i,p(j)}^H\mathbf{H}_{i,i,j}^H
            \clM_{i,j}^{p(j)}(\mathbf{V})^{-1}\mathbf{H}_{i,i,j}\mathbf{V}_{i, p(j)} \succeq \widetilde{\clQ}^{j,(\kappa)}_{i,p(j)}(\mathbf{V})           \label{lmi1} \qquad
\end{IEEEeqnarray}
for
\begin{IEEEeqnarray}{rCl}
 \widetilde{\mathbf{\clQ}}^{j,(\kappa)}_{i,p(j)}(\mathbf{V})&& \triangleq(\mathbf{V}^{(\kappa)}_{i,p(j)})^H\mathbf{H}_{i,i,j}^H\clM_{i,j}^{p(j)}(\mathbf{V}^{(\kappa)})^{-1}
\mathbf{H}_{i,i,j}\mathbf{V}_{i,p(j)} \nonumber\\
&&+\,(\mathbf{V}_{i,p(j)})^H\mathbf{H}_{i,i,j}^H\clM_{i,j}^{p(j)}(\mathbf{V}^{(\kappa)})^{-1}
\mathbf{H}_{i,i,j}\mathbf{V}^{(\kappa)}_{i,p(j)}\nonumber\\
&&-\,(\mathbf{V}^{(\kappa)}_{i,p(j)})^H\mathbf{H}_{i,i,j}^H\clM_{i,j}^{p(j)}(\mathbf{V}^{(\kappa)})^{-1}
\clM_{i,j}^{p(j)}(\mathbf{V})    \nonumber \\
&&\times \clM_{i,j}^{p(j)}(\mathbf{V}^{(\kappa)})^{-1}
\mathbf{H}_{i,i,j}\mathbf{V}^{(\kappa)}_{i,p(j)},\label{map1}
\end{IEEEeqnarray}
which also satisfies
\begin{IEEEeqnarray}{rCl}\label{match1}
&&\widetilde{\mathbf{\clQ}}^{j,(\kappa)}_{i,p(j)}(\mathbf{V}^{(\kappa)})= \nonumber\\
&&\qquad \; (\mathbf{V}_{i,p(j)}^{(\kappa)})^H\mathbf{H}_{i,i,j}^H
            \clM_{i,j}^{p(j)}(\mathbf{V}^{(\kappa)})^{-1}\mathbf{H}_{i,i,j}\mathbf{V}_{i, p(j)}^{(\kappa)}.\qquad
\end{IEEEeqnarray}
For
$\clM_{i,j}^{p(j)}(\mathbf{V})$ defined from (\ref{eq:Mijpj}), applying (\ref{in1}) again yields
\begin{eqnarray}
\clM_{i,j}^{p(j)}(\mathbf{V}) & \succeq &  \clL_{i,j}^{p(j),(\kappa)}(\mathbf{V})+\sigma^2\mathbf{I}_{N_r}\label{lmi1a}
\end{eqnarray}
over the trust region
\begin{equation}\label{tr1}
\clL_{i,j}^{p(j),(\kappa)}(\mathbf{V})\succeq  \mathbf{0},
\end{equation}
for the linear mapping
\begin{eqnarray}
 \clL_{i,j}^{p(j),(\kappa)}(\mathbf{V})\triangleq \ds\sum_{(s,l)\in{\cal S}\setminus\{(i,p(j))\}}\mathbf{H}_{s,i,j}
\Bigl[\mathbf{V}_{s,l}(\mathbf{V}_{s,l}^{(\kappa)})^H \nonumber\\
+\; \mathbf{V}_{s,l}^{(\kappa)}\mathbf{V}_{s,l}^H
-\mathbf{V}_{s,l}^{(\kappa)}(\mathbf{V}_{s,l}^{(\kappa)})^H\Bigl] \mathbf{H}_{s,i,j}^H.\label{linear1}
\end{eqnarray}
It follows from (\ref{lmi1}) that
\begin{equation}\label{lmi1.1}
\widetilde{\clQ}^{j,(\kappa)}_{i,p(j)}(\mathbf{V})\preceq \clQ^{j,(\kappa)}_{i,p(j)}(\mathbf{V})
\end{equation}
for
\begin{IEEEeqnarray}{rCl}
&&\clQ^{j,(\kappa)}_{i,p(j)}(\mathbf{V})\triangleq \mathbf{V}_{i,p(j)}^H\mathbf{H}_{i,i,j}^H\clM_{i,j}^{p(j)}(\mathbf{V}^{(\kappa)})^{-1}
\mathbf{H}_{i,i,j}\mathbf{V}^{(\kappa)}_{i,p(j)}\nonumber\\
&&\qquad+\, (\mathbf{V}^{(\kappa)}_{i,p(j)})^H\mathbf{H}_{i,i,j}^H\clM_{i,j}^{p(j)}(\mathbf{V}^{(\kappa)})^{-1}
\mathbf{H}_{i,i,j}\mathbf{V}_{i,p(j)}  \nonumber\\
&&\qquad-\,(\mathbf{V}^{(\kappa)}_{i,p(j)})^H\mathbf{H}_{i,i,j}^H\clM_{i,j}^{p(j)}(\mathbf{V}^{(\kappa)})^{-1}
\Bigl[\clL_{i,j}^{p(j),(\kappa)}(\mathbf{V})\nonumber \qquad \\
&&\qquad +\, \sigma^2\mathbf{I}_{N_r}\Bigr] \clM_{i,j}^{p(j)}(\mathbf{V}^{(\kappa)})^{-1}
\mathbf{H}_{i,i,j}\mathbf{V}^{(\kappa)}_{i,p(j)}.\label{map1:1}
\end{IEEEeqnarray}
Therefore,
\begin{IEEEeqnarray}{rCl}
&& \clR_{i,p(j)}^j(\mathbf{V})  \nonumber\\
                 &&= \ln\Bigl|\mathbf{I}_L + (\mathbf{V}_{i,p(j)})^H\mathbf{H}_{i,i,j}^H
            \clM_{i,j}^{p(j)}(\mathbf{V})^{-1}\mathbf{H}_{i,i,j}\mathbf{V}_{i, p(j)}\Bigl|\nonumber \quad\\
&&\geq\ln\Bigl|\mathbf{I}_L +\widetilde{\clQ}^{j,(\kappa)}_{i,p(j)}(\mathbf{V}) \Bigl|\label{lmi2}\\
&&\geq\clR_{i,p(j)}^j(\mathbf{V}^{(\kappa)})+L \nonumber\\
&&\quad -\; \Bigl\la \mathbf{I}_L +\widetilde{\clQ}^{j,(\kappa)}_{i,p(j)}(\mathbf{V}^{(\kappa)}),\Bigl(\mathbf{I}_L+ \widetilde{\clQ}^{j,(\kappa)}_{i,p(j)}(\mathbf{V})\Bigl)^{-1} \Bigr\ra\label{l1.0}  \\
&&\geq\widetilde{\clR}_{i,p(j)}^{j, (\kappa)}(\mathbf{V})\label{l1}
\end{IEEEeqnarray}
for
\begin{IEEEeqnarray}{rCl}
&&\widetilde{\clR}_{i,p(j)}^{j, (\kappa)}(\mathbf{V})
\triangleq\clR_{i,p(j)}^j(\mathbf{V}^{(\kappa)})+L  \nonumber\\
&&\qquad -\;  \Bigl\la \mathbf{I}_L +(\mathbf{V}_{i,p(j)}^{(\kappa)})^H\mathbf{H}_{i,i,j}^H
            \clM_{i,j}^{p(j)}(\mathbf{V}^{(\kappa)})^{-1}\nonumber \qquad\\
&&\qquad\qquad \times            \mathbf{H}_{i,i,j}\mathbf{V}_{i, p(j)}^{(\kappa)}
,\Bigl(\mathbf{I}_L+ \clQ^{j,(\kappa)}_{i,p(j)}(\mathbf{V})\Bigl)^{-1} \Bigr\ra,\label{l1.1}
\end{IEEEeqnarray}
which is a concave function. Inequality (\ref{lmi2}) follows from (\ref{lmi1}) and the condition
\[
\ln|\mathbf{I}_L+\mathbf{X}|\geq \ln|\mathbf{I}_L+\mathbf{Y}|,\ \forall\ \mathbf{X}\succeq\mathbf{Y}\succeq \mathbf{0}.
\]
Inequality (\ref{l1.0}) follows by applying (\ref{in2}) and using equality
\[
\ln\bigl|\mathbf{I}_L +\widetilde{\clQ}^{j,(\kappa)}_{i,p(j)}(\mathbf{V}^{(\kappa)}) \bigl|=
\clR_{i,p(j)}^j(\mathbf{V}^{(\kappa)}).
\]
Inequality (\ref{l1}) follows from inequality (\ref{lmi1}), equality (\ref{match1}),  and the condition
\[
\Bigl\la \mathbf{M},\mathbf{X}^{-1}\Bigl\ra \geq \Bigl\la \mathbf{M},\mathbf{Y}^{-1}\Bigl\ra, \ \forall\ \mathbf{M}\succeq \mathbf{0},
\mathbf{Y}\succeq \mathbf{X}\succ  \mathbf{0}.
\]
Analogously, the  rate function $\clR_{i,p(j)}^{p(j)}(\mathbf{V})$ and throughput function $\clR_{i,j}(\mathbf{V})$
defined from (\ref{eq:Rijj}) and (\ref{noma2}) are lower bounded by
\begin{IEEEeqnarray}{rCl}\label{eq:Ripjpj}
\clR_{i,p(j)}^{p(j)}(\mathbf{V})&\geq& \widetilde{\clR}_{i,p(j)}^{p(j), (\kappa)}(\mathbf{V})\label{l2}
\end{IEEEeqnarray}
over the trust region
\begin{equation}\label{tr2}
\clL_{i,p(j)}^{p(j),(\kappa)}(\mathbf{V})\succeq  \mathbf{0},
\end{equation}
and
\begin{IEEEeqnarray}{rCl}\label{eq:Rij}
\clR_{i,j}(\mathbf{V})&\geq&\widetilde{\clR}_{i,j}^{(\kappa)}(\mathbf{V})
\end{IEEEeqnarray}
over the trust region
\begin{equation}\label{tr3}
\clL_{i,j}^{p,(\kappa)}(\mathbf{V})\succeq  \mathbf{0},
\end{equation}
for linear mappings
\begin{IEEEeqnarray}{rCl}\label{linear2}
&&\clL_{i,p(j)}^{p(j),(\kappa)}(\mathbf{V}) \triangleq \ds\sum_{(s,l)\in{\cal S}\setminus\{(i,p(j)) \}}\mathbf{H}_{s,i,p(j)}\Bigl[\mathbf{V}_{s,l}(\mathbf{V}_{s,l}^{(\kappa)})^H  \nonumber\\
&&\qquad\qquad\quad +\; \mathbf{V}_{s,l}^{(\kappa)}\mathbf{V}_{s,l}^H-
\mathbf{V}_{s,l}^{(\kappa)}(\mathbf{V}_{s,l}^{(\kappa)})^H\Bigl]\mathbf{H}_{s,i,p(j)}^H
\end{IEEEeqnarray}
and
\begin{IEEEeqnarray}{rCl}\label{linear3}
&&\clL_{i,j}^{p,(\kappa)}(\mathbf{V}) \triangleq \ds\ds\sum_{(s,l)\in{\cal S}\setminus\{(i,p(j)),(i,j)\}}\mathbf{H}_{s,i,j}\Bigl[\mathbf{V}_{s,l}(\mathbf{V}_{s,l}^{(\kappa)})^H \nonumber\\
&&\qquad\qquad\quad +\; \mathbf{V}_{s,l}^{(\kappa)}\mathbf{V}_{s,l}^H-\mathbf{V}_{s,l}^{(\kappa)}(\mathbf{V}_{s,l}^{(\kappa)})^H\Bigl] \mathbf{H}_{s,i,j}^H,
\end{IEEEeqnarray}
and for concave functions
\begin{IEEEeqnarray}{rCl}
&&\widetilde{\clR}_{i,p(j)}^{p(j), (\kappa)}(\mathbf{V})
\triangleq \clR_{i,p(j)}^{p(j)}(\mathbf{V}^{(\kappa)})
   + L  \nonumber\\
&&\qquad -\Bigl\la \mathbf{I}_L +(\mathbf{V}^{(\kappa)}_{i,p(j)})^H\mathbf{H}_{i,i,p(j)}^H
            \clM_{i,p(j)}^{p(j)}(\mathbf{V}^{(\kappa)})^{-1}\nonumber\\
 &&\qquad \times           \mathbf{H}_{i,i,p(j)}\mathbf{V}_{i, p(j)}^{(\kappa)}
,\Bigl(\mathbf{I}_L +\clQ_{i,p(j)}^{p(j), (\kappa)}(\mathbf{V})\Bigl)^{-1}
\Bigr\ra\label{l2:2}
\end{IEEEeqnarray}
and
\begin{IEEEeqnarray}{rCl}
&&\widetilde{\clR}_{i,j}^{(\kappa)}(\mathbf{V}) \triangleq
\clR_{i,j}^{(\kappa)}(\mathbf{V}^{(\kappa)})
  + L \nonumber\\
&&\qquad - \; \Bigl\la \mathbf{I}_L +(\mathbf{V}^{(\kappa)}_{i,j})^H\mathbf{H}_{i,i,j}^H\clM_{i,j}^{p}(\mathbf{V}^{(\kappa)})^{-1}
\mathbf{H}_{i,i,j}\mathbf{V}_{i,j}^{(\kappa)}, \nonumber\\
&& \qquad\qquad \Bigl(\mathbf{I}_L +\clQ^{(\kappa)}_{i,j}(\mathbf{V})\Bigl)^{-1}\Bigr\ra\label{l3},
\end{IEEEeqnarray}
with
\begin{IEEEeqnarray}{rCl}
&&\clQ_{i,p(j)}^{p(j), (\kappa)}(\mathbf{V})\triangleq  \nonumber\\
&&\quad   (\mathbf{V}^{(\kappa)}_{i,p(j)})^H\mathbf{H}_{i,i,p(j)}^H
            \clM_{i,p(j)}^{p(j)}(\mathbf{V}^{(\kappa)})^{-1}\mathbf{H}_{i,i,p(j)}\mathbf{V}_{i, p(j)} \nonumber\\
&&\quad +\, (\mathbf{V}_{i,p(j)})^H\mathbf{H}_{i,i,p(j)}^H
            \clM_{i,p(j)}^{p(j)}(\mathbf{V}^{(\kappa)})^{-1}\mathbf{H}_{i,i,p(j)}\mathbf{V}^{(\kappa)}_{i, p(j)} \nonumber\\
&&\quad -\, (\mathbf{V}^{(\kappa)}_{i,p(j)})^H\mathbf{H}_{i,i,p(j)}^H
            \clM_{i,p(j)}^{p(j)}(\mathbf{V}^{(\kappa)})^{-1}
            \Bigl[\clL_{i,p(j)}^{p(j),(\kappa)}(\mathbf{V})  \nonumber \quad \\
&&\quad +\; \sigma^2\mathbf{I}_{N_r}\Bigr]
            \clM_{i,p(j)}^{p(j)}(\mathbf{V}^{(\kappa)})^{-1}
            \mathbf{H}_{i,i,p(j)}\mathbf{V}^{(\kappa)}_{i, p(j)}          \label{map2}\quad\
\end{IEEEeqnarray}
and
\begin{IEEEeqnarray}{rCl}
\clQ^{(\kappa)}_{i,j}(\mathbf{V})&\triangleq& (\mathbf{V}^{(\kappa)}_{i,j})^H\mathbf{H}_{i,i,j}^H\clM_{i,j}^{p}(\mathbf{V}^{(\kappa)})^{-1}
\mathbf{H}_{i,i,j}\mathbf{V}_{i,j}   \nonumber\\
&+&\;(\mathbf{V}_{i,j})^H\mathbf{H}_{i,i,j}^H\clM_{i,j}^{p}(\mathbf{V}^{(\kappa)})^{-1}
\mathbf{H}_{i,i,j}\mathbf{V}^{(\kappa)}_{i,j}\nonumber\\
&-&\,(\mathbf{V}^{(\kappa)}_{i,j})^H\mathbf{H}_{i,i,j}^H\clM_{i,j}^{p}(\mathbf{V}^{(\kappa)})^{-1}
\Bigl[\clL_{i,j}^{p,(\kappa)}(\mathbf{V}) \nonumber\\
&+&\; \sigma^2\mathbf{I}_{N_r}\Bigr]\clM_{i,j}^{p}(\mathbf{V}^{(\kappa)})^{-1}
\mathbf{H}_{i,i,j}\mathbf{V}^{(\kappa)}_{i,j}.\label{map3}
\end{IEEEeqnarray}
We  also define
\begin{equation}\label{l4}
\widetilde{\clR}^{(\kappa)}_{i,p(j)}(\mathbf{V})\triangleq \min\Bigl\{\widetilde{\clR}^{j,(\kappa)}_{i,p(j)}(\mathbf{V}),
\widetilde{\clR}^{p(j),(\kappa)}_{i,p(j)}(\mathbf{V})\Bigl\}.
\end{equation}

In summary, at the $\kappa$-th iteration, the following SDP, which is an inner approximation of \eqref{Rmtgor}, is solved
to generate the next feasible point $\mathbf{V}^{(\kappa+1)}$:
     \begin{IEEEeqnarray}{rCl}\label{sdp1}
   &&     \ds\underset{\mathbf{V}}{\max} \ \, \mathcal{\widetilde{P}}^{(\kappa)}(\mathbf{V}) \triangleq\sum_{i=1}^N\sum_{j=1}^K \Bigl(\widetilde{\clR}_{i,j}^{(\kappa)}(\mathbf{V}) + \widetilde{\clR}^{(\kappa)}_{i,p(j)}(\mathbf{V}) \Bigr)\IEEEyessubnumber\label{sdp1:a}\qquad\\
 &&  \st \
 \widetilde{\clR}_{i,j}^{(\kappa)}(\mathbf{V})\geq r_{i,j},\ \forall i\in\mathcal{I},\ \forall j\in\mathcal{K}_1,\IEEEyessubnumber\label{sdp1:b}\\
  && \qquad\qquad\quad   \widetilde{\clR}^{(\kappa)}_{i,p(j)}(\mathbf{V}) \geq r_{i,p(j)},  \; \forall i\in\mathcal{I},\;  \forall j\in\mathcal{K}_2,\IEEEyessubnumber\label{sdp1:c}\\
 &&\qquad\qquad\quad \eqref{Rmtgord}, \eqref{tr1}, \eqref{tr2}, \eqref{tr3}.\IEEEyessubnumber\label{sdp1:d}
        \end{IEEEeqnarray}
The proposed Algorithm~\ref{alg_SDP_FW}
generates a sequence $\{\bV^{(\kappa)}\}$ of improved points of  \eqref{Rmtgor}, which also converges to a KKT point.
			
\begin{algorithm}[t]
\begin{algorithmic}[1]
\protect\caption{SDP-based path-following algorithm for the STM \eqref{Rmtgor} in MIMO-NOMA}
\label{alg_SDP_FW}
\global\long\def\algorithmicrequire{\textbf{Initialization:}}
\REQUIRE  Set $\kappa:=0$ and solve \eqref{kiter2}  to generate an initial feasible point $\mathbf{V}^{(0)}$ for constraints
 \eqref{Rmtgorb}-\eqref{Rmtgord}.
\REPEAT
\STATE Solve the semi-definite program \eqref{sdp1} to obtain the optimal solution: $\mathbf{V}^{\star}$.
\STATE Update\ \ $\mathbf{V}^{(\kappa+1)}:= \mathbf{V}^{\star}.$
\STATE Set $\kappa:=\kappa+1.$
\UNTIL Convergence\\
\end{algorithmic} \end{algorithm}			
				
A feasible point $\mathbf{V}^{(0)}$ for the constraints \eqref{Rmtgorb}-\eqref{Rmtgord} to initialize Algorithm \ref{alg_SDP_FW}
can be found by invoking the iterations
\begin{IEEEeqnarray}{rCl}\label{kiter2}
&&\max_{\mathbf{V}} \underset{(i,j)\in\mathcal{S}}{\min}\min \Biggl\{\frac{\widetilde{\clR}^{(\kappa)}_{i,j}(\mathbf{V})}{r_{i,j}},
\frac{\widetilde{\clR}^{(\kappa)}_{i,p(j)}(\mathbf{V})}{r_{i,p(j)}}\Biggl\}\ \IEEEyessubnumber\\
&&\st\quad \eqref{Rmtgord},
\eqref{tr1}, \eqref{tr2}, \eqref{tr3} \IEEEyessubnumber
\end{IEEEeqnarray}
to reach a value more than or equal to $1$ in satisfying \eqref{Rmtgorb}-\eqref{Rmtgord}.

\textit{Complexity analysis:} The SDP \eqref{sdp1} involves $N(1+3K)$ quadratic constraints, $3NK$ semi-definite
constraints with $N_r$ rows and $n=KN(2N_tL+1)$ real decision variables. For $m_{\text{SDP}}\triangleq N(1+3K)+3NKN_r$,
its computational complexity is $\mathcal{O}(n^2m_{\text{SDP}}^{2.5}+m_{\text{SDP}}^{3.5})$, which is seen higher than that of
the convex quadratic problem \eqref{RmtgFWA}.

\section{Tailored algorithm for MISO-NOMA}\label{sec:TL}
In this case, all channels are row vectors ($\bH_{s,i,j}\in\mathbb{C}^{1\times N_t}$)
and $s_{i,j}\in\mathbb{C}$ ($L=1$).

As observed first time in \cite{WES06}, for
\begin{equation}\label{varch}
\bar{\bV}_{i,j}=e^{-\jmath {\sf arg}(\bH_{i,i,j}\bV_{i,j})}\bV_{i,j}
\end{equation}
one has
$|\bH_{i,i,j}\bV_{i,j}|=\bH_{i,i,j}\bar{\bV}_{i,j}=\Re\{ \bH_{i,i,j}\bar{\bV}_{i,j}\}\geq 0$ and
$|\bH_{i',i,j'}\bV_{i,j}|=|\bH_{i',i,j'}\bar{\bV}_{i,j}|$ for $(i',j')\neq (i,j)$. Therefore,
without loss of generality we can replace
\[
\bH_{i,i,j}\bV_{i,j}\bV_{i,j}^H\bH_{i,i,j}^H = \bigl|\bH_{i,i,j}\bV_{i,j}\bigr|^2, \ j=1,\dots, 2K
\]
by
\[
\bigl(\Re\{\bH_{i,i,j}\bV_{i,j}\}\bigr)^2, \  j=1,\dots, 2K
\]
with
\begin{equation}\label{positivereal}
\Re\{\bH_{i,i,j}\bV_{i,j}\}\geq 0, \ j=1,\dots, 2K
\end{equation}
(including $p(j)$ for $j=K+1,...,2K$).
Accordingly, write
\[
\clM_{i,j}(\mathbf{V})= \ds\sum_{(s,l)\in{\cal S}}\bigl|\mathbf{H}_{s,i,j}\mathbf{V}_{s,l}\bigr|^2+ \sigma^2.
\]
Then the message $\mathbf{s}_{i,p(j)}$ intended for cell-edge UE $(i,p(j))$ is decoded by the cell-center  UE $(i,j)$
with the achievable rate
\begin{equation}\label{t1}
r_{i,p(j)}^j(\mathbf{V}) =  \ln\Biggl(1 + \frac{\bigl|\mathbf{H}_{i,i,j}\mathbf{V}_{i,p(j)}\bigr|^2}{\clM_{i,j}^{p(j)}(\mathbf{V})}\Biggr),
        \end{equation}
and is decoded by the cell-edge UE $(i,p(j))$ itself  with the achievable rate
       \begin{equation}\label{t2}
            r_{i,p(j)}^{p(j)}(\mathbf{V}) = \ln\Biggl( 1 +
           \frac{\bigl(\Re\{ \mathbf{H}_{i,i,p(j)}\mathbf{V}_{i, p(j)}\}\bigr)^2}{\clM_{i,p(j)}^{p(j)}(\mathbf{V})} \Biggl)
        \end{equation}
where
\begin{IEEEeqnarray}{rCl}
\clM_{i,j}^{p(j)}(\mathbf{V}) & \triangleq&  \clM_{i,j}(\mathbf{V}) - \bigl|\mathbf{H}_{i,i,j}\mathbf{V}_{i,p(j)}\bigr|^2 \nonumber\\
&=&\ds\ds\sum_{(s,l)\in{\cal S}\setminus\{(i,p(j))\}}
\bigl|\mathbf{H}_{s,i,j}\mathbf{V}_{s,l}\bigl|^2 + \sigma^2, \nonumber
\end{IEEEeqnarray}
and
\begin{IEEEeqnarray}{rCl}
\clM_{i,p(j)}^{p(j)}(\mathbf{V})& \triangleq& \clM_{i,p(j)}(\mathbf{V}) - \bigl|\mathbf{H}_{i,i,p(j)}\mathbf{V}_{i,p(j)}\bigl|^2 \nonumber\\
&=&\ds\sum_{(s,l)\in{\cal S}\setminus\{(i,p(j)) \}}
\bigl|\mathbf{H}_{s,i,p(j)}\mathbf{V}_{s,l}\bigl|^2 + \sigma^2.\nonumber
\end{IEEEeqnarray}
Also, the message $\mathbf{s}_{i,j}$ intended for the cell-center UE $(i,j)$ is successively decoded by UE $(i,j)$ itself
with the throughput
\begin{equation}\label{t3}
            r_{i,j}(\mathbf{V}) = \ln\Biggr( 1 + \frac{\bigl(\Re\{\mathbf{H}_{i,i,j}\mathbf{V}_{i,j}\}\bigr)^2}{ \clM_{i,j}^{p}(\mathbf{V})} \Biggr)
 \end{equation}
        where
       \begin{IEEEeqnarray}{rCl}
        \clM_{i,j}^{p}(\mathbf{V}) & \triangleq & \clM_{i,j}^{p(j)}(\mathbf{V}) - \bigl|\mathbf{H}_{i,i,j}\mathbf{V}_{i,j}\bigl|^2 \nonumber\\
        &=&\ds\ds\sum_{(s,l)\in{\cal S}\setminus\{(i,p(j)),(i,j)\}}\bigl|\mathbf{H}_{s,i,j}\mathbf{V}_{s,l}\bigl|^2 + \sigma^2. \nonumber\nonumber
        \end{IEEEeqnarray}

For $r_{i,p(j)}(\mathbf{V}) \triangleq \min\bigl\{r^j_{i,p(j)}(\mathbf{V}), r^{p(j)}_{i,p(j)}(\mathbf{V})\bigl\}$, the problem (\ref{Rmtgor}) in this case is
\begin{IEEEeqnarray}{rCl}\label{beam1}
        &&\underset{\mathbf{V}}{\maxi}\ \mathcal{P}(\mathbf{V})\triangleq\sum_{i=1}^N\sum_{j=1}^K\Bigr(r_{i,j}(\mathbf{V})+ r_{i,p(j)}(\mathbf{V})
            \Bigl)\quad \IEEEyessubnumber\label{beam1a} \quad\\
				&& \st \
        r_{i,j}(\mathbf{V}) \geq r_{i,j},  \ \forall i\in\mathcal{I},\ \forall j\in\mathcal{K}_1, \IEEEyessubnumber\label{beam1b} \\
        &&\qquad\qquad\   r^{p(j)}_{i,p(j)}(\mathbf{V}) \geq r_{i,p(j)},  \ \forall i\in\mathcal{I},\ \forall j\in\mathcal{K}_2, \IEEEyessubnumber\label{beam1c}\\
        &&\qquad\qquad\  r^{j}_{i,p(j)}(\mathbf{V}) \geq r_{i,p(j)},  \ \forall i\in\mathcal{I},\ \forall j\in\mathcal{K}_2, \IEEEyessubnumber\label{beam1d}\\
        &&\qquad\qquad\  \sum_{j\in \clJ} \|\mathbf{V}_{i,j}\|^2  \le P^{\max}_{i}, \forall i \in\mathcal{I}. \IEEEyessubnumber\label{beam1e}
\end{IEEEeqnarray}

Due to the above transforms (\ref{t2}) and (\ref{t3}) under condition (\ref{positivereal}), the nonconvex
constraints (\ref{beam1b}) and (\ref{beam1c}) are expressed by the second-order cone (SOC) constraints
\begin{IEEEeqnarray}{rCl}
\Re\bigl\{\mathbf{H}_{i,i,j}\mathbf{V}_{i,j}\bigr\} &\geq& \sqrt{e^{r_{i,j}}-1}\sqrt{\clM_{i,j}^{p}(\mathbf{V})}, \nonumber\\
&& \ \forall i\in\mathcal{I},\ \forall j\in\mathcal{K}_1, \label{beam1be} \\
 \Re\bigl\{ \mathbf{H}_{i,i,p(j)}\mathbf{V}_{i, p(j)} \bigl\} &\geq&
 \sqrt{e^{r_{i,p(j)}}-1}\sqrt{\clM_{i,p(j)}^{p(j)}(\mathbf{V})},  \qquad \nonumber\\
&&   \ \forall i\in\mathcal{I},\ \forall j\in\mathcal{K}_2\label{beam1ce}
\end{IEEEeqnarray}
but the constraint (\ref{beam1d}) remains to be nonconvex.

To approximate functions in (\ref{beam1a}) we use the inequality
\begin{equation}\label{zf8}
\ln(1+z)\geq a(\bar{z})-b(\bar{z})\frac{1}{z},\quad
\forall\ z>0, \bar{z}>0
\end{equation}
with
\begin{equation}\label{zf9}
0<a(\bar{z}) \triangleq \ln(1+\bar{z})+\ds\frac{\bar{z}}{\bar{z}+1},\
0<b(\bar{z}) \triangleq \frac{\bar{z}^2}{\bar{z}+1},
\end{equation}
whose proof is provided by Appendix~C.

Applying (\ref{zf8}) for
 $\bar{z}=z^{j, (\kappa)}_{i,p(j)}\triangleq |\mathbf{H}_{i,i,j}\mathbf{V}^{(\kappa)}_{i,p(j)}|^2/\clM_{i,j}^{p(j)}(\mathbf{V}^{(\kappa)})$ and
 $z= |\mathbf{H}_{i,i,j}\mathbf{V}_{i,p(j)}|^2/\clM_{i,j}^{p(j)}(\mathbf{V})$ yields
\begin{IEEEeqnarray}{rCl}
&&r^{j}_{i,p(j)}(\mathbf{V})\geq a\bigl(z_{i,p(j)}^{j,(\kappa)}\bigl)-b\bigl(z_{i,p(j)}^{j,(\kappa)}\bigl)\ds\frac{\clM_{i,j}^{p(j)}(\mathbf{V})}{\bigl|\mathbf{H}_{i,i,j}\mathbf{V}_{i,p(j)}\bigr|^2} \nonumber\\
&&\qquad\qquad\ \geq  r^{j, (\kappa)}_{i,p(j)}(\mathbf{V}) \nonumber\\
&&\qquad\qquad\ \triangleq a\bigl(z_{i,p(j)}^{j,(\kappa)}\bigl)-
b\bigl(z_{i,p(j)}^{j,(\kappa)}\bigl)
 \ds\frac{\clM_{i,j}^{p(j)}(\mathbf{V})}{\varphi_{i,p(j)}^{j,(\kappa)}(\mathbf{V})} \qquad
\end{IEEEeqnarray}
over the trust region
\begin{IEEEeqnarray}{rCl}\label{trb1}
\varphi_{i,p(j)}^{j,(\kappa)}(\mathbf{V}) \triangleq
2\Re\bigl\{\mathbf{H}_{i,i,j}\mathbf{V}^{(\kappa)}_{i,p(j)}
(\mathbf{H}_{i,i,j}\mathbf{V}_{i,p(j)})^*\bigr\} && \nonumber\\
 -\ \bigl|\mathbf{H}_{i,i,j}\mathbf{V}^{(\kappa)}_{i,p(j)}\bigl|^2 &&\; > 0.\qquad
\end{IEEEeqnarray}
Analogously,
\begin{IEEEeqnarray}{rCl}
r^{p(j)}_{i,p(j)}(\mathbf{V})&\geq&r^{p(j), (\kappa)}_{i,p(j)}(\mathbf{V}) \nonumber\\
&\triangleq&a\bigl(z_{i,p(j)}^{p(j),(\kappa)}\bigl) -
b\bigl(z_{i,p(j)}^{p(j),(\kappa)}\bigl)\ds\frac{\clM_{i,p(j)}^{p(j)}(\mathbf{V})}
{\varphi_{i,p(j)}^{p(j),(\kappa)}(\mathbf{V})}\qquad
\end{IEEEeqnarray}
with
\begin{IEEEeqnarray}{rCl}
\varphi_{i,p(j)}^{p(j),(\kappa)}(\mathbf{V}) \triangleq&& \Re\bigl\{\mathbf{H}_{i,i,p(j)}\mathbf{V}^{(\kappa)}_{i,p(j)}\bigl\}
\Bigl(2\Re\bigl\{\mathbf{H}_{i,i,p(j)}\mathbf{V}_{i,p(j)}\bigl\}  \nonumber \\
 && -\ \Re\bigl\{\mathbf{H}_{i,i,p(j)}\mathbf{V}^{(\kappa)}_{i,p(j)}\bigl\}\Bigl) \nonumber
\end{IEEEeqnarray}
over the trust region
\begin{equation}\label{trb2}
2\Re\bigl\{\mathbf{H}_{i,i,p(j)}\mathbf{V}_{i,p(j)}\bigl\}-\Re\bigl\{\mathbf{H}_{i,i,p(j)}\mathbf{V}^{(\kappa)}_{i,p(j)}\bigl\}\; > 0,
\end{equation}
and
\begin{IEEEeqnarray}{rCl}
r_{i,j}(\mathbf{V})&\geq& r^{(\kappa)}_{i,p(j)}(\mathbf{V}) \nonumber\\
&\triangleq&a(z_{i,j}^{(\kappa)})-
b(z_{i,j}^{(\kappa)})\ds\frac{\clM_{i,j}^{p}(\mathbf{V})}{ \varphi_{i,j}^{(\kappa)}(\mathbf{V})}
\end{IEEEeqnarray}
with
\begin{IEEEeqnarray}{rCl}
\varphi_{i,j}^{(\kappa)}(\mathbf{V}) \triangleq&&\ \Re\bigl\{\mathbf{H}_{i,i,j}\mathbf{V}^{(\kappa)}_{i,j}\bigl\}
\Bigl(2\Re\bigl\{\mathbf{H}_{i,i,j}\mathbf{V}_{i,j}\bigl\} \nonumber\\
&&-\ \Re\bigl\{\mathbf{H}_{i,i,j}\mathbf{V}^{(\kappa)}_{i,j}\bigl\}\Bigl) \nonumber
\end{IEEEeqnarray}
over the trust region
\begin{equation}\label{trb3}
2\Re\bigl\{\mathbf{H}_{i,i,j}\mathbf{V}_{i,j}\bigl\}-\Re\bigl\{\mathbf{H}_{i,i,j}\mathbf{V}^{(\kappa)}_{i,j}\bigl\} > 0.
\end{equation}
In Algorithm \ref{alg_Tailored_FW}, we propose an QP-based path-following algorithm to solve problem (\ref{beam1}). At
the $\kappa$-th iteration it solves the following SOC program to generate the next feasible point $\mathbf{V}^{(\kappa+1)}$:
   \begin{IEEEeqnarray}{rCl}\label{beamk}
      &&    \underset{\mathbf{V}}{\max} \quad \mathcal{P}^{(\kappa)}(\mathbf{V}) \triangleq\sum_{i=1}^N\sum_{j=1}^K \Bigl(r_{i,j}^{(\kappa)}(\mathbf{V})+r_{i,p(j)}^{(\kappa)}(\mathbf{V})  \Bigl)          \IEEEyessubnumber \qquad\\
        && \st\  r^{j,(\kappa)}_{i,p(j)}(\mathbf{V})\geq r_{i,p(j)}, \ \forall i\in\mathcal{I},\ \forall j\in\mathcal{K}_2, \IEEEyessubnumber \\
       &&\qquad\qquad\   \eqref{beam1e}, (\ref{beam1be}), (\ref{beam1ce}), (\ref{trb1}), (\ref{trb2}), (\ref{trb3})  \IEEEyessubnumber
   \end{IEEEeqnarray}
where $r_{i,p(j)}^{(\kappa)}(\mathbf{V}) \triangleq \min\Bigl\{r^{j, (\kappa)}_{i,p(j)}(\mathbf{V}), r^{p(j), (\kappa)}_{i,p(j)}(\mathbf{V})\Bigr\}$.				
				
\begin{algorithm}[t]
\caption{Tailored QP-based path-following algorithm  for the
STM \eqref{beam1} in MISO-NOMA}\label{alg_Tailored_FW}
\begin{algorithmic}
\STATE \textbf{Initialization}: Initialize a feasible point $\mathbf{V}^{(0)}$ for constraints in (\ref{beam1}).
\STATE \textbf{$\kappa$-th iteration}: Solve the convex quadratic program (\ref{beamk}) to find the optimal solution $\mathbf{V}^{\star}$. If $\bigr|\bigl({\cal P}(\mathbf{V}^{\star}) - \mathcal{P}(\mathbf{V}^{(\kappa)})\bigr) \big/ \mathcal{P}(\mathbf{V}^{(\kappa)}) \bigl| \le \epsilon$, terminate. Otherwise, set $\kappa :=\kappa+1, \mathbf{V}^{(\kappa)}:=\mathbf{V}^{\star}$ and continue.
\end{algorithmic}
\end{algorithm}
To find a feasible point for constraints in (\ref{beam1}) for initializing Algorithm \ref{alg_Tailored_FW},
initialized by a feasible point $\mathbf{V}^{(0)}$ for
the convex constraints (\ref{beam1e}), (\ref{beam1be}), and (\ref{beam1ce}), the following SOC based iterations
\begin{IEEEeqnarray}{rCl}\label{beami}
&&\max_{\mathbf{V}} \min_{(i,j)\in\mathcal{S}_2}\ \Biggl\{\frac{r^{j,(\kappa)}_{i,p(j)}
(\mathbf{V})}{r_{i,p(j)}}\Biggr\}  \quad  \IEEEyessubnumber\\
&&\st\quad
\eqref{beam1e}, (\ref{beam1be}), (\ref{beam1ce}), (\ref{trb1})\IEEEyessubnumber
\end{IEEEeqnarray}
are invoked for reaching a value more or equal to $1$ in satisfying constraints in (\ref{beam1}).

\textit{Complexity analysis:} The SOC program \eqref{beamk} involves $m_{\text{SOC}}=N(1+6K)$ quadratic or
SOC constraints and $n=KN(2N_tL+1)$ real decision variables. Its computational complexity is $\mathcal{O}(n^2m_{\text{SOC}}^{2.5}+m_{\text{SOC}}^{3.5})$, which is seen higher than that of
the convex quadratic problem \eqref{RmtgFWA}.

\section{Numerical Results}\label{sec:simulation}
\begin{figure}[t]
\centering
\includegraphics[width=0.45\textwidth,trim={0cm 0.0cm -0cm -0cm}]{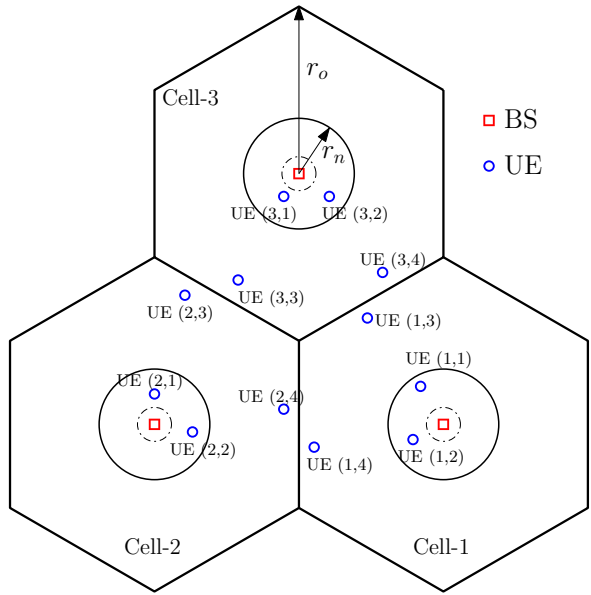}
\caption{MIMO-NOMA multi-cell system.}
\label{fig:SM:RatePerUE}
\end{figure}

In this section we use numerical examples to evaluate the performance of the proposed algorithms.
A system topology shown in Fig. \ref{fig:SM:RatePerUE} is set up. There are
$N=3$ macro cells and $4$  UEs per cell with two cell-center UEs and two cell-edge UEs, which are
located near to the boundaries with the two adjacent cells. Unless stated otherwise, $N_t = 4$ and $N_r = 2$
are set for MIMO-NOMA, for which $L=N_r$ is set. Thus, the precoder-matrices $\mathbf{V}_{i,j}$ are of dimension $N_t\times N_r$.
The channel matrix between a BS and a UE at a  distance $d$ (in kilometres) is generated  as $\mathbf{H}=\sqrt{10^{-\sigma_{\mathsf{PL}}/10}}\tilde{\mathbf{H}}$ \cite{ETSI}. Here, $\sigma_{\mathsf{PL}}$ is the path loss (PL) in dB and the entries of $\tilde{\mathbf{H}}$ are independent and identically distributed complex Gaussian variables with zero mean and unit variance. Without loss of generality,  the requirement  thresholds for
all UEs are set as $r_{i,j} = r_{i,p(j)}\equiv \bar{\mathsf{R}}$ and  the same
power budget $P^{\max}_i = P^{\max},\, \forall i\in\mathcal{I}$ is given to all BSs.

 For the ease of reference, the other parameters given in Table \ref{parameter} including  $\bar{\mathsf{R}}$
 are used.  The  error tolerance in the proposed Algorithms is set to  $\epsilon=10^{-3}$. The numerical results are obtained using the parser YALMIP \cite{L04}. The achieved sum throughput results are divided by $\ln(2)$ to arrive at the unit of bps/channel-use.
Each simulation is run $100$ times and the result are averaged to arrive at the final figures.

\begin{table}[t]
\caption{SIMULATION PARAMETERS}
	\label{parameter}
	\centering
	{\setlength{\tabcolsep}{0.695em}
\setlength{\extrarowheight}{0.6em}
		\begin{tabular}{l|l}
		\hline
				Parameters & Value \\
		\hline\hline
		    Carrier frequency/ Bandwidth                             & 2 [GHz]/ 20 [MHz] \\
				Noise power density & -174 [dBm/Hz] \\
				Path loss from the  BS to a  UE, $\sigma_{\mathsf{PL}}$   & 128.1 + 37.6$\log_{10}(d)$ [dB]\\
				Shadowing standard deviation & 8 [dB] \\
				Radius of each cell, $r_o$ &  500 [m]\\
				Coverage of near UEs, $r_n$  & 150 [m]\\
				Distance between BS and nearest UE & $>$ 10 [m]\\
				Threshold  $\bar{\mathsf{R}}$   & 1 [bps/Hz]  \\
		\hline		   				
		\end{tabular}}
\end{table}	

\subsection{Algorithms' Convergence}

\begin{figure}
    \begin{center}
    \begin{subfigure}[MIMO-NOMA networks.]{
        \includegraphics[width=0.49\textwidth]{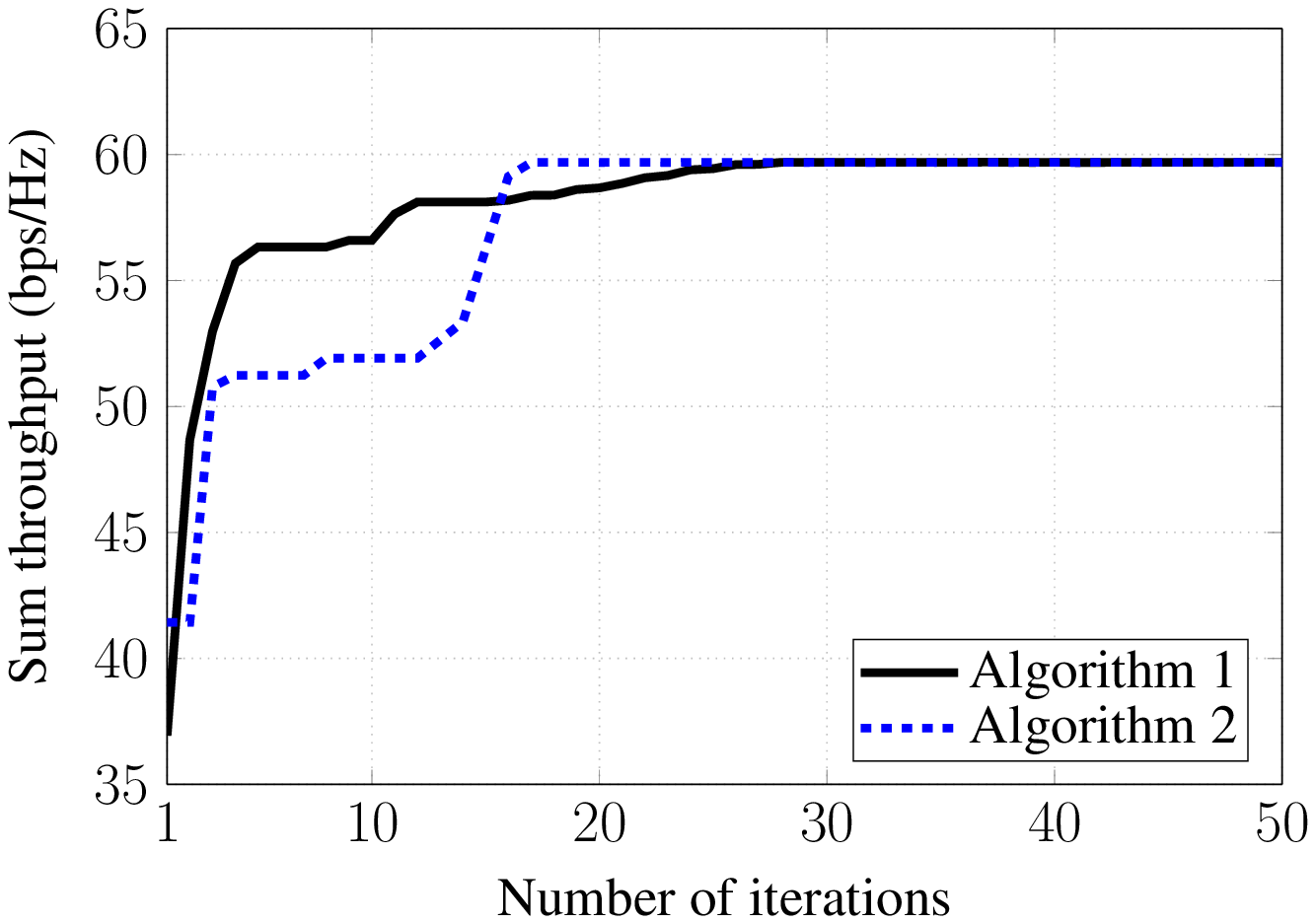}}
    		\label{fig:a}
				\end{subfigure}
				 \begin{subfigure}[MISO-NOMA networks.]{
        \includegraphics[width=0.49\textwidth]{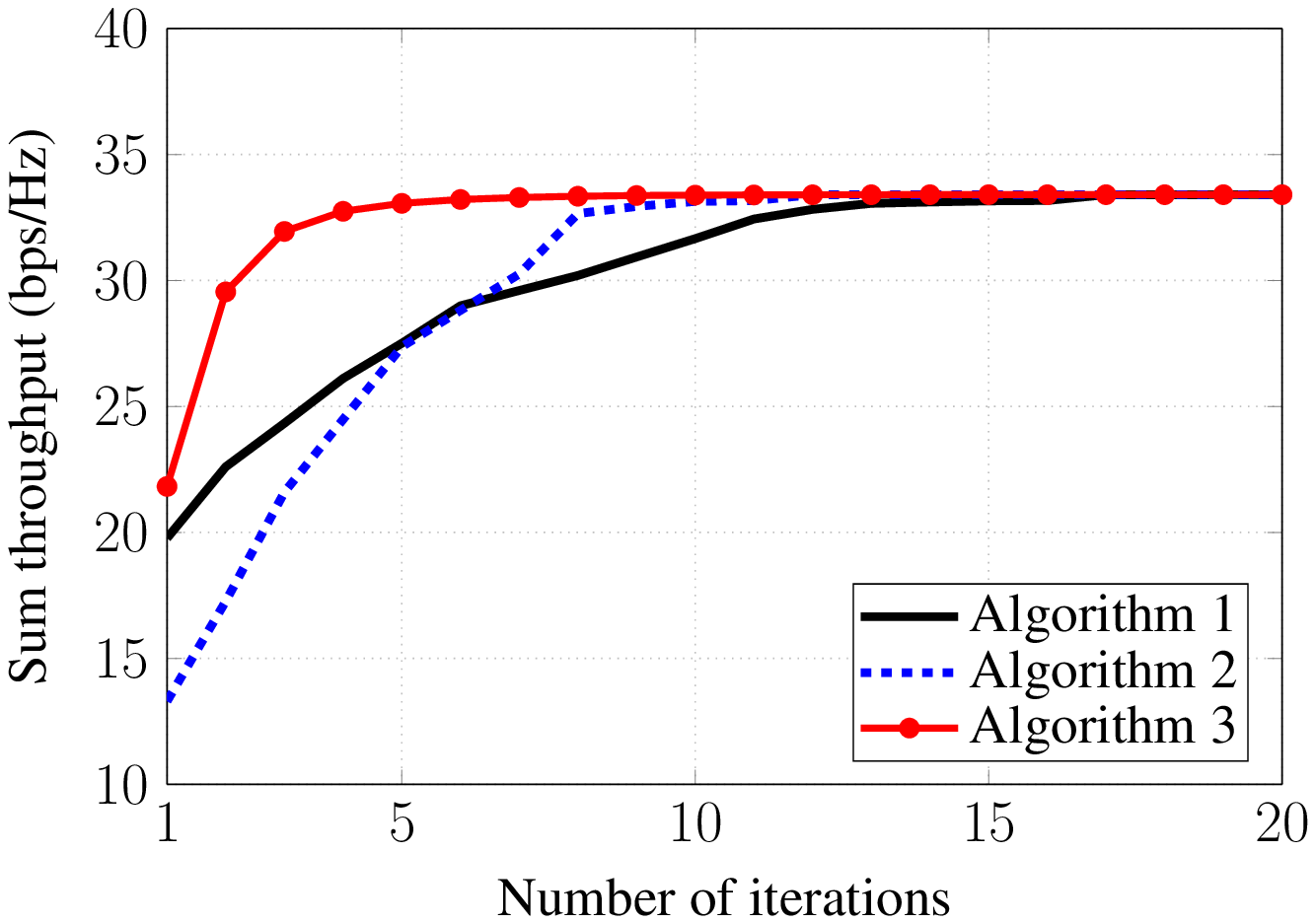}}
        \label{fig:b}
    \end{subfigure}
	  \caption{Convergence pattern for $P^{\max} = 30$ dBm.}\label{fig:Convergencebehavior:Iteration}
\end{center}
\end{figure}

Fig.~\ref{fig:Convergencebehavior:Iteration} shows the typical convergence behavior of the proposed algorithms for a given set of channel realizations that are randomly generated  for the two cases.
According to Fig.~\ref{fig:Convergencebehavior:Iteration}(a), both Algorithm~\ref{alg_SCALE_FW} and \ref{alg_SDP_FW}
 for MIMO-NOMA reach the almost optimal value of sum throughput in  $18$ and $13$ iterations.
 As expected, Algorithm~\ref{alg_SDP_FW} converges faster than Algorithm~\ref{alg_SCALE_FW}.
 On the other hand, according to Fig.~\ref{fig:Convergencebehavior:Iteration}(b),  Algorithm~\ref{alg_Tailored_FW} for NOMA-MISO
 converges very fast reaching the optimal value in 6 iterations.

\subsection{Numerical Results for MIMO-NOMA}

\begin{figure}
    \begin{center}
    \begin{subfigure}[Sum throughput versus  $P^{\max}$.]{
        \includegraphics[width=0.48\textwidth]{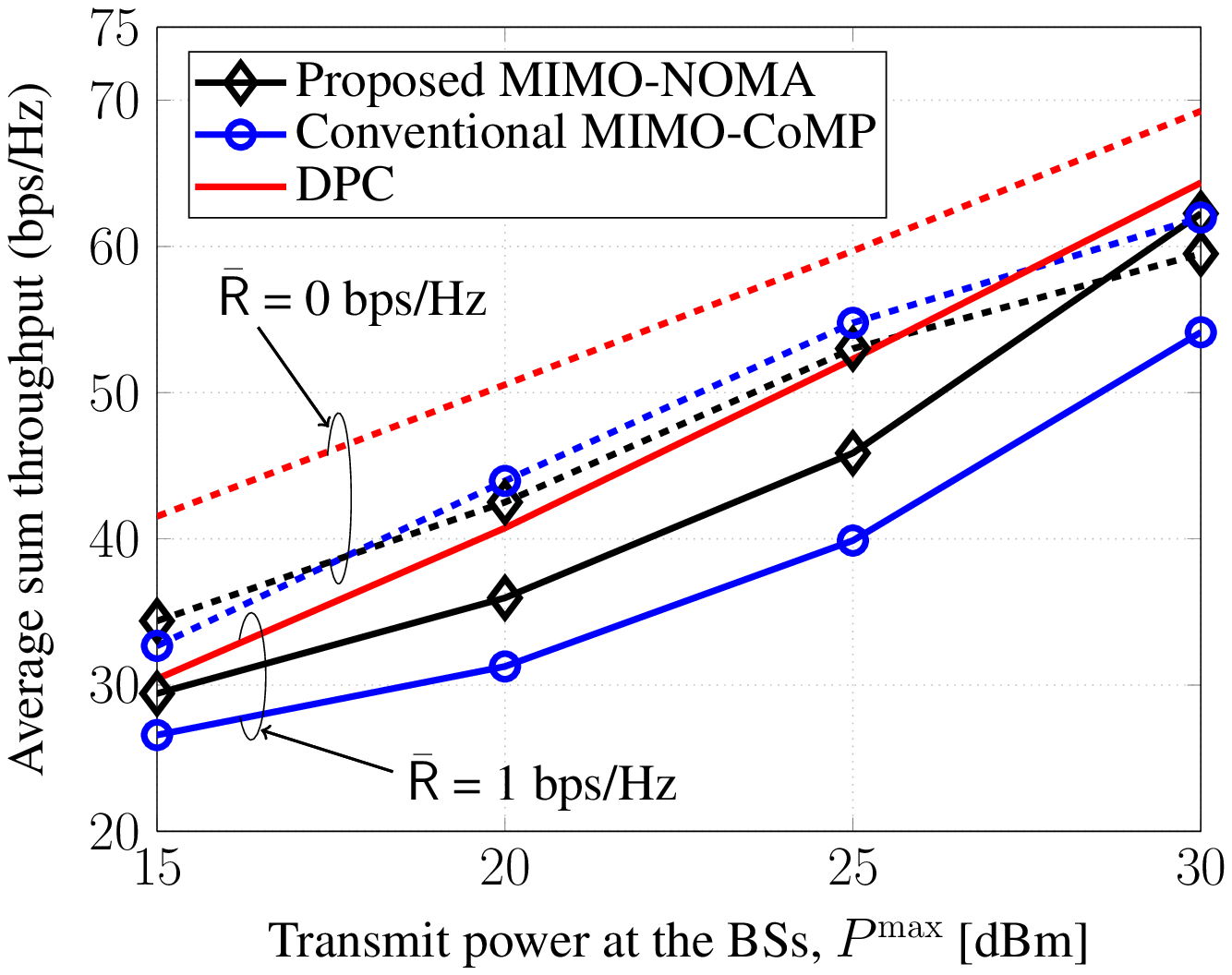}}
    		\label{fig:SRMMIMO:Pmax}
				\end{subfigure}
				 \begin{subfigure}[Sum throughput versus $\bar{\mathsf{R}}$ with $P^{\max} = 30$ dBm.]{
        \includegraphics[width=0.48\textwidth]{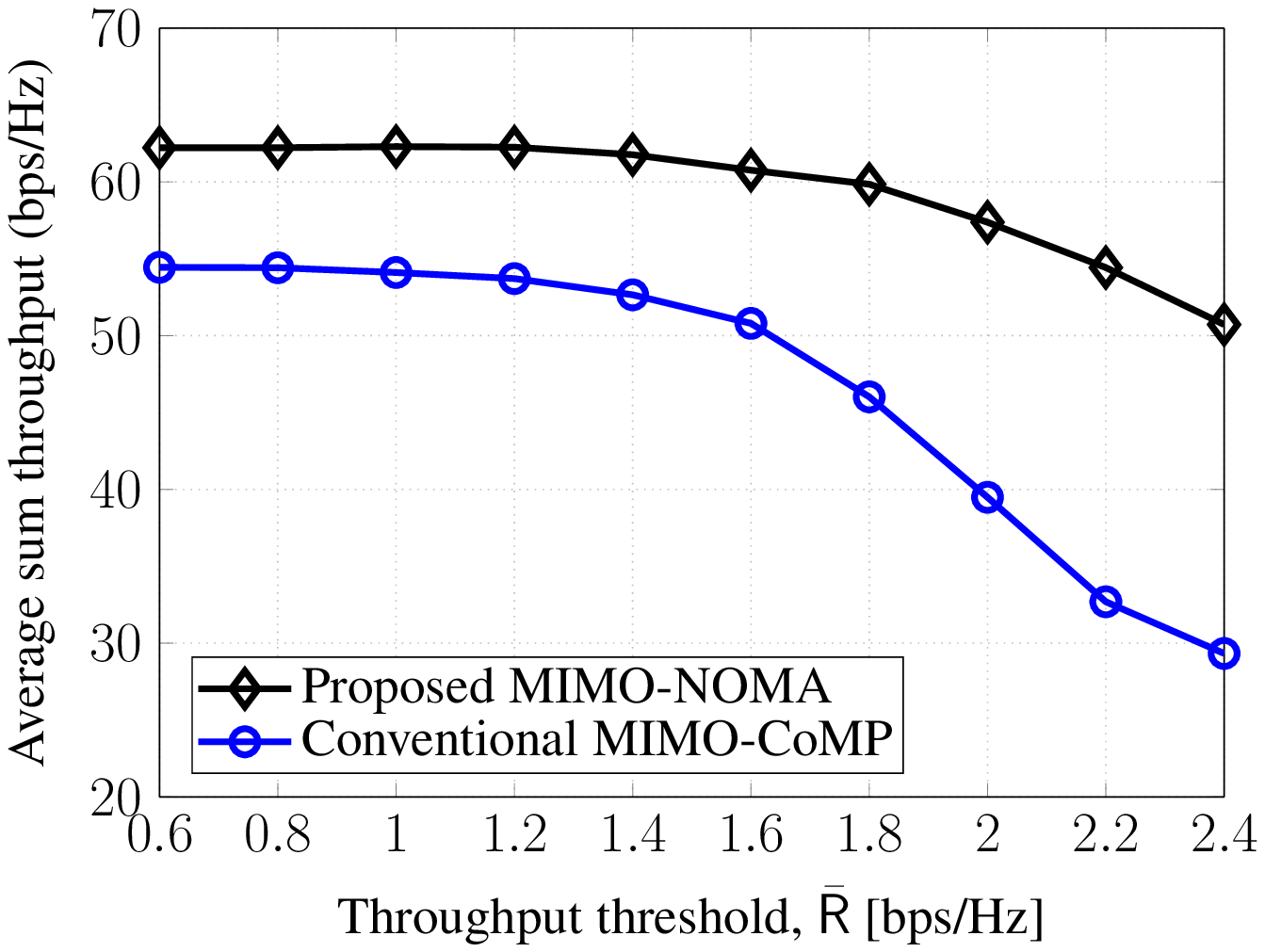}}
        \label{fig:SRMMIMO:Rbar}
    \end{subfigure}
	  \caption{Sum throughput of MIMO-NOMA, (a) versus the transmit power at the BSs and (b) versus the throughput threshold ($N = 3, K = 2, N_t = 4$, and $N_r = 2$).}\label{fig:SRMMIMO:PmaxRbar}
\end{center}
\end{figure}

\begin{table}[t]
\centering
\caption{Achieved User Throughput (bps/Hz) in Multi-user MIMO Multi-cell Systems for $\bar{\mathsf{R}} = 0$ bps/Hz}
\label{Table:RateperUENoQoS}
{\setlength{\tabcolsep}{0.50em}
\setlength{\extrarowheight}{0.65em}
\begin{tabular}{|c|c|c|c|c|c|c|}
\hline
\multicolumn{1}{|c|}{\multirow{2}{*}{}} & \multicolumn{3}{c|}{Throughput per UE} & \multicolumn{3}{c|}{ST per cell}                                                 \\ \cline{2-7}
\multicolumn{1}{|c|}{}                  & NOMA        & CoMP        & DPC        & \multicolumn{1}{c|}{NOMA} & \multicolumn{1}{c|}{CoMP} & \multicolumn{1}{c|}{DPC} \\ \hline
UE (1,1)                                & 6.36        & 8.27        & 9.43       & \multirow{4}{*}{18.48}    & \multirow{4}{*}{19.11}    & \multirow{4}{*}{22.54}   \\ \cline{1-4}
UE (1,2)                                & 7.13        & 9.39        & 7.86       &                           &                           &                          \\ \cline{1-4}
UE (1,3)                                & 1.39        & 1.07        & 3.27       &                           &                           &                          \\ \cline{1-4}
UE (1,4)                                & 3.60        & 0.38        & 1.98       &                           &                           &                          \\ \hline
UE (2,1)                                & 5.96        & 9.70        & 10.76      & \multirow{4}{*}{18.50}    & \multirow{4}{*}{23.35}    & \multirow{4}{*}{24.01}   \\ \cline{1-4}
UE (2,2)                                & 7.02        & 7.87        & 8.93       &                           &                           &                          \\ \cline{1-4}
UE (2,3)                                & 3.59        & 2.19        & 2.69       &                           &                           &                          \\ \cline{1-4}
UE (2,4)                                & 1.93        & 3.59        & 1.63       &                           &                           &                          \\ \hline
UE (3,1)                                & 5.04        & 7.62        & 10.06      & \multirow{4}{*}{21.27}    & \multirow{4}{*}{19.25}    & \multirow{4}{*}{22.62}   \\ \cline{1-4}
UE (3,2)                                & 7.30        & 8.44        & 8.24       &                           &                           &                          \\ \cline{1-4}
UE (3,3)                                & 4.40        & 0.63        & 2.86       &                           &                           &                          \\ \cline{1-4}
UE (3,4)                                & 4.53        & 2.56        & 1.46       &                           &                           &                          \\ \hline
\multicolumn{4}{|c|}{Total ST}                                                   & \underline{\textbf{58.25}}                     & \underline{\textbf{61.71}}                     & \underline{\textbf{69.17}}                    \\ \hline
\end{tabular}}
\end{table}

For the ease of reference, the result achieved by \eqref{CoMPobj} is labeled by ``Conventional MIMO-CoMP'' whereas
that achieved by \eqref{Rmtgor} is labeled by  ``Proposed MIMO-NOMA.''  Fig.~\ref{fig:SRMMIMO:PmaxRbar}(a) plots
the sum throughput  versus the power budget $P^{\max}$ under setting $\bar{\sf R} = \{0,1\} $ bps/Hz.
For $\bar{\sf R}=0$ bps/Hz, i.e. there is no UEs' QoS requirement imposed, CoMP slightly outperforms
MIMO-NOMA by achieving throughput concentrated at the cell-center UEs of good conditions.
Table~\ref{Table:RateperUENoQoS} details the UEs's throughput distribution for $P^{\max}=30$ dBm.
The high ratio $9.39/0.38=24.7$ between the best UE throughput and the worst UE throughput (BWR) implies that CoMP would
perform wobbly in the QoS maximization problem (\ref{qos}) (with $r_{i,j}\equiv 1$), which also expresses the
system ability to offer the uniform service to UEs. BWR for MIMO-NOMA is $5.1=7.13/1.39$ so it is expected to outperform
CoMP in maximizing (\ref{qos}).  The low throughput $1.46$ bps/Hz
at UE $(3,4)$ by DPC is a result of a strong interference from an adjacent cell, which cannot be mitigated by DPC.

For $\bar{\sf R}=1$ bps/Hz, MIMO-NOMA of course offers a higher sum throughput than
CoMP, where the BSs are seen spending a nearly full power budget
$P^{\max}$  in gaining the sum throughput. Increasing $P^{\max}$ also leads to a remarkable gain
  in sum throughput by NOMA compared with  CoMP. The sum throughput by the former also  catches up
 that by DPC.   The gain of MIMO-NOMA is  a result of canceling interference from intra-cluster interference, as shown in \eqref{eq:Mipjpj}. The cell-center UEs in CoMP experience intra-cluster interference that  becomes stronger  when transmit power increases.
   The plot of sum throughput versus QoS requirement threshold $\bar{\mathsf{R}}\in[0.6,\,2.4]$ bps/Hz is shown by Fig.~\ref{fig:SRMMIMO:PmaxRbar}(b) for $P^{\max} = 30$ dBm. The sum throughput are nearly flat for   $\bar{\mathsf{R}} \leq 1.2$ bps/Hz and are degraded after that.
    The BSs in CoMP must allocate much more power to serve cell-edge UEs when QoS threshold increases. As a result,
    the system sum throughput is dropped quickly.
     In contrast, the sum throughput of MIMO-NOMA is still slightly sensitive to QoS requirement threshold
     because BSs can tune the  power allocation in meeting cell-edge UEs' QoS requirements
     whenever the cell-center UEs' QoS requirement is  easily met.

\begin{figure}
    \begin{center}
    \begin{subfigure}[Sum throughput versus  $K$ with $N_t = 4$.]{
        \includegraphics[width=0.48\textwidth]{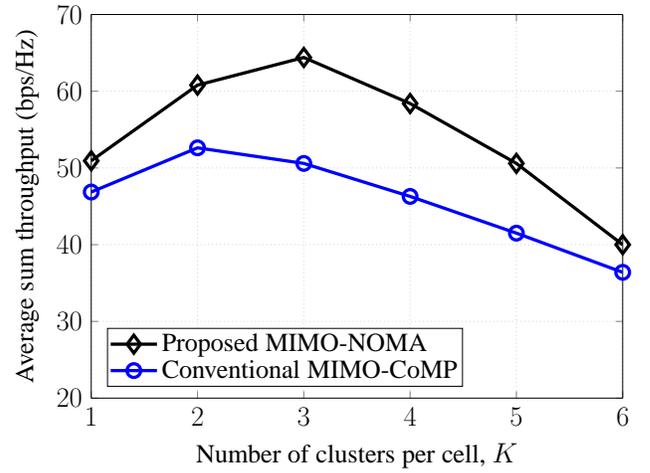}}
    		\label{fig:SRMMIMO:K}
				\end{subfigure}
				 \begin{subfigure}[Sum throughput versus $N_t$ with $K = 2$.]{
        \includegraphics[width=0.48\textwidth]{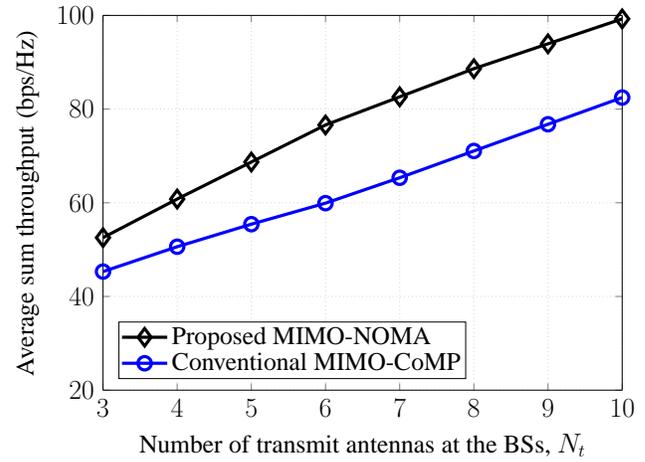}}
        \label{fig:SRMMIMO:N}
    \end{subfigure}
	  \caption{Sum throughput of the MIMO-NOMA, (a) versus the number of clusters per cell and (b) versus the number of transit antennas at the BS ($N = 3, N_r = 2$, and $P^{\max} = 30$ dBm).}\label{fig:SRMMIMO:KN}
\end{center}
\end{figure}

Fig.~\ref{fig:SRMMIMO:KN} shows the impact of the number of  UEs per cell and the number of transmit antennas at the BS on the performance of the system. Fig.~\ref{fig:SRMMIMO:KN}(a) shows that MIMO-NOMA can deliver an
acceptable sum throughput for large $K$. Again, MIMO-NOMA outperforms CoMP in all $K$.
The sum throughput of the both systems  decreases  from a certain value of $K$
where there is not much degree-of-freedom (DoF) for leveraging multi-user diversity. Interestingly, MIMO-NOMA achieves its best
sum throughput for $K=3$ (6 UEs) while  CoMP is peaked at $K=2$ (4 UEs).
Of course, these numbers are not magic and  can be changed in other settings.  Fig.~\ref{fig:SRMMIMO:KN}(b)
plots the sum throughput vs the number $N_t$ of antennas at the BSs.

\begin{table}[t]
\centering
\caption{Achieved User Throughput (bps/Hz) in Multi-user MIMO Multi-cell Systems}
\label{Table:RateperUE}
{\setlength{\tabcolsep}{0.45em}
\setlength{\extrarowheight}{0.65em}
\begin{tabular}{|c|c|c|c|c|c|c|}
\hline
\multirow{2}{*}{} & \multicolumn{2}{c|}{Throughput per UE} & \multicolumn{2}{c|}{ST per cell}                 & \multicolumn{2}{c|}{Total ST}                   \\ \cline{2-7}
                &  CoMP         &   NOMA        &           CoMP         &             NOMA       &      CoMP               &NOMA                     \\ \hline
  UE (1,1)   &8.37&7.83& \multirow{4}{*}{20.68} & \multirow{4}{*}{27.16} & \multirow{12}{*}{\underline{\textbf{59.63}}} & \multirow{12}{*}{\underline{\textbf{77.23}}} \\ \cline{1-3}
  UE (1,2)   &8.74&7.62&                   &                   &                    &                    \\ \cline{1-3}
  UE (1,3)   &2.32&5.12&                   &                   &                    &                    \\ \cline{1-3}
  UE (1,4)   &1.25&6.59&                   &                   &                    &                    \\ \cline{1-5}
  UE (2,1)   &7.81&7.56& \multirow{4}{*}{18.76} & \multirow{4}{*}{24.34} &                    &                    \\ \cline{1-3}
  UE (2,2)   &6.31&6.24&                   &                   &                    &                    \\ \cline{1-3}
  UE (2,3)   &3.13&5.78&                   &                   &                    &                    \\ \cline{1-3}
  UE (2,4)   &1.51&4.76&                   &                   &                    &                    \\ \cline{1-5}
  UE (3,1)   &9.01&8.86& \multirow{4}{*}{20.19} & \multirow{4}{*}{25.73} &                    &                    \\ \cline{1-3}
  UE (3,2)   &6.25&6.92&                   &                   &                    &                    \\ \cline{1-3}
  UE (3,3)   &1.89&6.33&                   &                   &                    &                    \\ \cline{1-3}
  UE (3,4)   &3.04&3.62&                   &                   &                    &                    \\ \hline
\end{tabular}}
\end{table}

Table~\ref{Table:RateperUE} details the throughput at UEs under setting $N_t=6$ and $P^{\max}=30$ dBm.
Under the same QoS requirement threshold for UEs, the CoMP's throughput  is mostly contributed
by the cell-center UEs, i.e. CoMP still tends to punish  the cell-edge UEs, who are in poor channel condition.
Raising the QoS requirement for the cell-edge UEs to counter this discrimination would lead to the risk of
CoMP service feasibility. In contrast, in maximizing the system's throughput, MIMO-NOMA offers
much fairer and balanced services without contrasting the QoS requirement thresholds so it is very suitable for new
quality-of-experience (QoE) services for cell-edge UEs.
\begin{figure}
    \begin{center}
		 \begin{subfigure}[CDF  versus sum throughput for $P^{\max} = 15$ dBm.]{
        \includegraphics[width=0.48\textwidth]{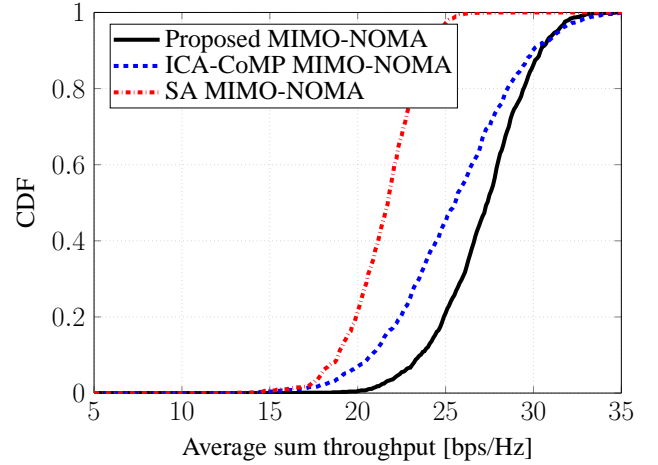}}
        \label{fig:CDF:b}
    \end{subfigure}
				\begin{subfigure}[CDF versus sum throughput  for $P^{\max} = 30$ dBm.]{
        \includegraphics[width=0.48\textwidth]{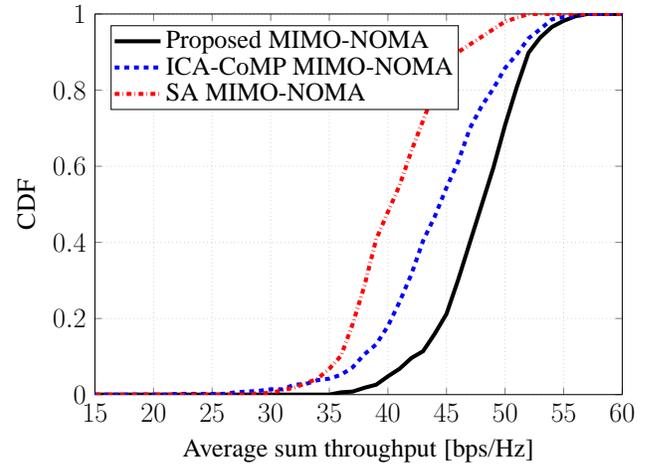}}
    		\label{fig:CDF:a}
				\end{subfigure}
	  \caption{CDF of the sum throughput (a) for $P^{\max} = 30$ dBm and (b) for $P^{\max} = 15$ dBm with $N = 2, K = 4, N_t = 5$, and $ N_r = 4$.}\label{fig:CDFMIMO}
\end{center}
\end{figure}

By Fig.~\ref{fig:CDFMIMO},
the system's throughput achieved by the proposed precoder design  is compared to that achieved by
signal alignment MIMO-NOMA (SA MIMO-NOMA) \cite{DSP16} and interfering channel alignment CoMP MIMO-NOMA (ICA-CoMP MIMO-NOMA) \cite{ShinComL16} in the two-cell scenario  with $K = 4$ users, $N_t = 5$, while  $ N_r = 4$ is set to make the signal
and channel alignment feasible.  In SA MIMO-NOMA, the inter-cluster  interference is canceled
by detection vectors based SA technique at the UEs and zero-forcing (ZF) based precoder matrix at the BSs. In ICA-CoMP MIMO-NOMA,
a receive beamformer is constructed at the cell-edge UEs to align the  interfering channels, and then a transmit beamformer  based on the null space at the BS is designed to ensure zero inter-cell and inter-cluster interference. For the cell-center UEs,  a ZF decoder is designed to cancel the inter-cluster interference only. To ensure a fair comparison, it is additionally
assumed that the cell-center UEs do not experience inter-cell interference as
they are far away from the   neighboring cell in practice \cite{ShinComL16}.
The two different settings of $P^{\max}$  = (15 dBm, 30 dBm), according to 3GPP TR 36.942 v.9.0.1 with a 46-dBm
maximum transmit power for the 20 MHz bandwidth are under consideration.
Fig.~\ref{fig:CDFMIMO} plots the cumulative distribution function (CDF) of the sum throughput. As expected,  MIMO-NOMA  outperforms both ICA-CoMP MIMO-NOMA and SA MIMO-NOMA. Specifically,
by controlling the interference to the cell-edge UEs more efficiently, MIMO-NOMA  reaches
1.5 bps/Hz and 5.9 bps/Hz  higher than the ICA-CoMP MIMO-NOMA and SA MIMO-NOMA, respectively, in about $60\%$ of the simulated trials with $P^{\max}$  = 15 dBm (see Fig.~\ref{fig:CDFMIMO}(a)). With $P^{\max}$  = 30 dBm, it is even more essential (see Fig.~\ref{fig:CDFMIMO}(b)).

\subsection{Numerical Results for MISO-NOMA}

\begin{figure}[t]
\centering
\includegraphics[width=0.48\textwidth,trim={0cm 0.0cm -0cm -0cm}]{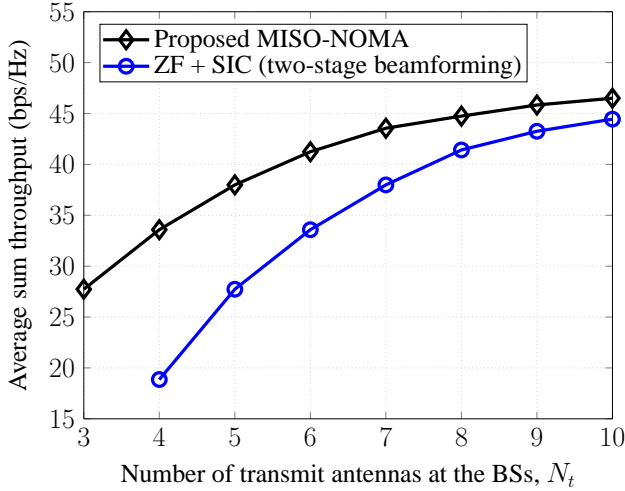}
\caption{Average sum throughput of the MISO-NOMA  versus the number of transit antennas at the BS ($N = 3, K = 2$, and $P^{\max} = 30$ {dBm}).}
\label{fig:SRMMISO:N}
\end{figure}

\begin{figure}[t]
\centering
\includegraphics[width=0.48\textwidth,trim={0cm 0.0cm -0cm -0cm}]{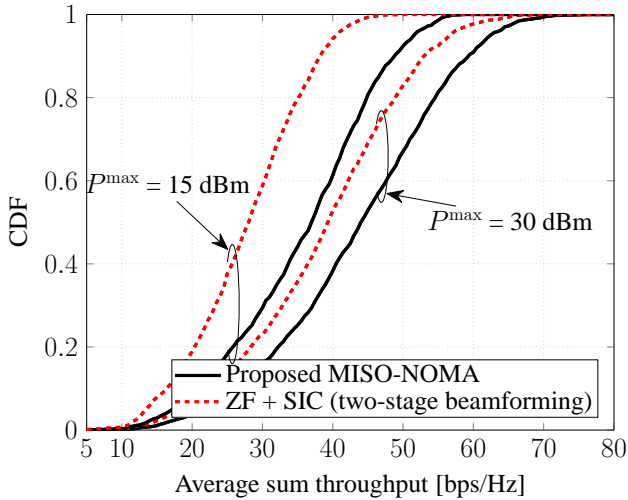}
\caption{CDF of the sum throughput for  two different settings of the transmit power ($N = 3, K = 2, N_t = 6$, and $ N_r = 1$).}
\label{fig:SRMMISO:CDF}
\end{figure}

The performance of MISO-NOMA achieved by \eqref{beam1} is compared  to that achieved
by the two-stage beamforming in \cite{Choi15}. The key idea of  the two-stage beamforming
is that  ZF beamforming is employed at the BS first to cancel the inter-pair interference and
then SIC is used for each pair of UEs. The optimal solution for two-stage beamforming  can be easily found by using Algorithm~\ref{alg_Tailored_FW}.

Fig.~\ref{fig:SRMMISO:N} plots the total sum throughput vs. the number of transmit antennas at the BS
under setting $N=3$, $K=2$ and $P^{\max} = 30$ dBm. Note that two-stage beamforming requires  $N_t\geq 2(K-1) + 2=4$.
Unsurprisingly, MISO-NOMA achieves a better sum throughput than the two-stage beamforming.
Two-stage beamforming achieves closer performance to MISO-NOMA  as the number $N_t$
of transmit antennas increases, providing  more degrees of freedom to leverage multi-user diversity.
Fig.~\ref{fig:SRMMISO:CDF} plots the CDF of the sum throughput at $N_t = 6$.
The performance gap is narrower at a higher power budget.
\subsection{Comparison for Different Cluster Sizes}

\begin{figure}
    \begin{center}
		 \begin{subfigure}[Simulation setup considered in Fig.~\ref{fig:SRM3UE}(b) and Fig.~\ref{fig:SRM3UE}(c) with $r_n = 100 $ m, $r_m = 250 $ m, and $r_o = 500 $ m.]{
        \includegraphics[width=0.35\textwidth]{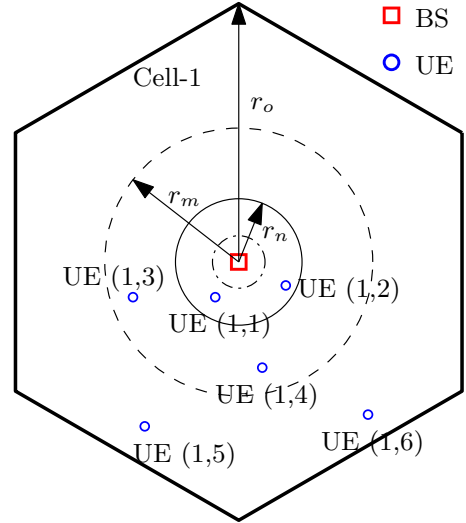}}
        \label{fig:SRM3UE:a}
    \end{subfigure}
				\begin{subfigure}[Average sum throughput of the MIMO-NOMA  versus $\bar{\mathsf{R}}$ for $N_r = 2$.]{
        \includegraphics[width=0.47\textwidth]{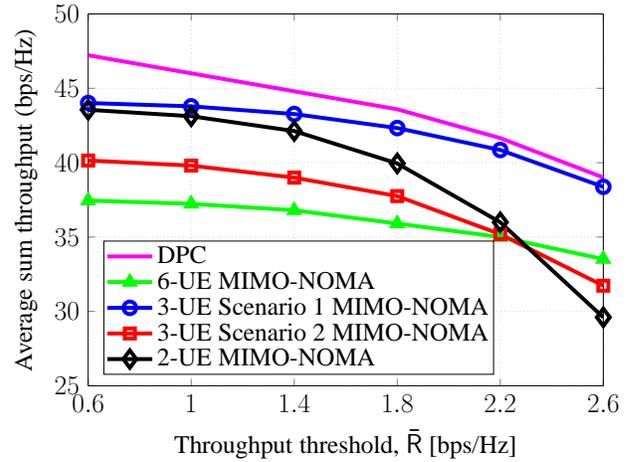}}
    		\label{fig:SRM3UE:b}
				\end{subfigure}
				\begin{subfigure}[Average sum throughput of the MISO-NOMA  versus $\bar{\mathsf{R}}$ for $N_r = 1$.]{
        \includegraphics[width=0.47\textwidth]{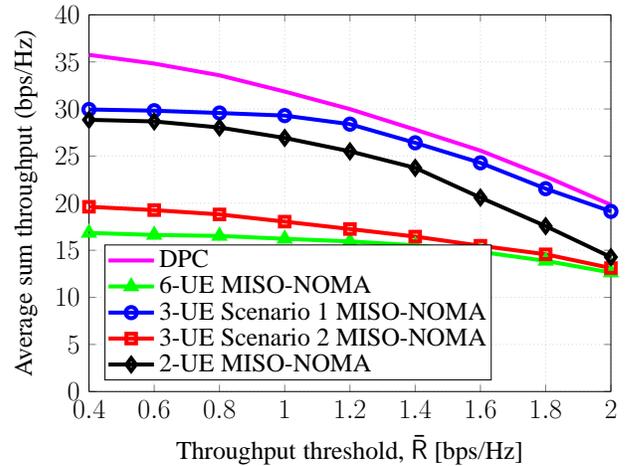}}
    		\label{fig:SRM3UE:c}
				\end{subfigure}
	  \caption{Average sum throughput (b) for the MIMO-NOMA, and (c) for the MISO-NOMA  with different cluster sizes ($N_t = 12$ and $P^{\max} = 30$ dBm).}\label{fig:SRM3UE}
\end{center}
\end{figure}

 As the last numerical example, we investigate the system performance
in a single-cell scenario of more than two UEs grouped to create a virtual cluster for NOMA. There are 6 UEs in total,
which are randomly placed in three different areas, as shown by Fig.~\ref{fig:SRM3UE}(a). Two cell-center  UEs
are located inside the disc of radius $r_n$ = 100 m, two cell-middle UEs
 are located inside the ring  of inner radius $100$ m and outer radius $250$ m, and
  two cell-edge UEs  are located inside the ring of inner radius $250$ m and outer radius $500$ m.
 Different cluster sizes are considered: two UEs per cluster, three UEs per cluster, and six UEs per cluster.
 For two-UE-per-cluster, a cell-center UE is randomly paired with a cell-middle UE, while the other
  is randomly paired with a cell-edge UE. The unpaired cell-middle UE and cell-edge are then paired to create the third cluster.
There are two scenarios in grouping for three-UE-per-cluster.
\begin{itemize}
   \item scenario-1 (\textit{more distinct channel conditions}): each cluster consists of a cell-center UE, a cell-middle UE
   and a cell-edge UE;
	  \item scenario-2 (\textit{less distinct channel conditions}): each cluster consists of a cell-center UE.
\end{itemize}
The order of decoding messages for UEs in the same cluster of size three is as follows:
the message for a cell-edge UE is decoded by all UEs, the message for a cell-middle UE is decoded by itself and the third
UE by canceling the previously decoded message for the cell-edge UE from the intra-cluster,
and the message  for the third UE is decoded by itself only by canceling all previously decoded messages from the
intra-cluster interference. Analogously, the  messages for UEs $(1,1)$, $(1,2)$, $(1,3)$, $(1,4)$, $(1,5)$ and $(1,6)$
are successively decoded and canceled from the intra-cell interference in the case of six-UE-per-cluster, i.e. all UEs
of the same cell have NOMA.
The proposed algorithms are easily adapted  for solution of the corresponding sum throughput maximization problems.

Fig.~\ref{fig:SRM3UE}(b) and Fig.~\ref{fig:SRM3UE}(c) plot the sum throughput
achieved by different clustering versus throughput threshold for  MIMO-NOMA and MISO-NOMA
under setting $N_t = 12$ and $P^{\max} = 30$ dBm.
 In general, the sum throughput achieved by the three-UE-per-cluster and six-UE-per-cluster based  schemes are
  dropped less than that achieved by two-UE-per-cluster based one when the threshold
   $\bar{\mathsf{R}}$ raises. Specifically, in Fig.~\ref{fig:SRM3UE}(b),  the sum throughput
    of the three-UE scenario 2 and six-UE are worse than the two-UE  for $\bar{\mathsf{R}} < 2.4$ bps/Hz and vice versa. The BS will allocate a much higher transmit power to the UE of the worst channel condition
     in the three-UE and six-UE schemes  than in the two-UE scheme to meet the QoS constraints. In other words, the cell-edge UEs' throughput  is significantly improved in larger cluster sizes.

   Notably, the sum throughput by three-UE scenario 1 catches up that by   DPC
    for larger $\bar{\mathsf{R}}$ in both Fig.~\ref{fig:SRM3UE}(b) and  Fig.~\ref{fig:SRM3UE}(c).
   In addition, six-UE scheme cannot provide a good sum throughput since large UEs per cluster may have error propagation in SIC leading to drastically reduce NOMA performance.  Another interesting observation
is that the sum throughput of the three-UE scenario 1 outperforms that of the other schemes.  Recalling that NOMA is more efficient by exploiting their  channel condition differences, i.e. 3-UE scenario 1 with more distinct channel conditions.  Consequently, a larger cluster size is recommended for more distinct channel conditions while a smaller cluster size is recommended for less distinct channel conditions.

\section{Conclusions}\label{sec:conclusion}
We have addressed the problem of sum throughput maximization in NOMA based systems
by proposing new path-following optimization algorithms. Numerical examples with realistic parameters have confirmed
 their fast convergence to an optimal solution. They reveal that NOMA not only  helps
 increase the cell-edge UEs' throughput substantially but also achieves much higher total sum throughput.
The appropriate size of UE cluster  with more distinctive channel gains has also been  shown to achieve
 remarkable  gains in NOMA systems.
\section*{Appendix A: Proof for inequality (\ref{in1})}
By \cite[Appendix B]{RTKN14}
\begin{IEEEeqnarray}{rCl}
\bigl(\alpha \mathbf{V}_1+\beta\mathbf{V}_2\bigr)^H\bigl(\alpha\mathbf{X}_1+\beta\mathbf{X}_2\bigl)^{-1}\bigl(\alpha \mathbf{V}_1+\beta\mathbf{V}_2\bigl)
\preceq&& \nonumber\qquad\\
\alpha \mathbf{V}_1^H\mathbf{X}_1^{-1}\mathbf{V}_1+
\beta\mathbf{V}_2^H\mathbf{X}_2^{-1}\mathbf{V}_2&&
\end{IEEEeqnarray}
for all $\alpha\geq 0$, $\beta\geq 0$, $\alpha+\beta=1$ and $\mathbf{V}_1$, $\mathbf{V}_2$, $\mathbf{X}_1\succ  \mathbf{0} $,
$\mathbf{X}_2\succ  \mathbf{0} $.  This means for all $\mathbf{x}$, function
\begin{IEEEeqnarray}{rCl}
f(\mathbf{V},\mathbf{X})=\mathbf{x}^H\mathbf{V}^H\mathbf{X}^{-1}\mathbf{V}\mathbf{x}
\end{IEEEeqnarray}
is convex. Then, for all $\mathbf{V}$, $\bar{\mathbf{V}}$, $\mathbf{X}\succ  \mathbf{0} $,
$\bar{\mathbf{X}}\succ  \mathbf{0} $ it is true that \cite{Tuybook}
\begin{IEEEeqnarray}{rCl}
f(\mathbf{V},\mathbf{X})&\geq& f(\bar{\mathbf{V}},\bar{\mathbf{X}}) + \bigl\la \nabla f(\bar{\mathbf{V}},\bar{\mathbf{X}}),
(\mathbf{V},\mathbf{X})-(\bar{\mathbf{V}},\bar{\mathbf{X}})\bigl\ra \nonumber\\
&=&\mathbf{x}^H\Bigl[\bar{\boldsymbol{V}}^H\bar{\boldsymbol{X}}^{-1}\mathbf{V}+
\mathbf{V}^H\bar{\boldsymbol{X}}^{-1}\bar{\boldsymbol{V}}  \nonumber\\
&&\qquad\quad  -\; \bar{\boldsymbol{V}}^H\bar{\boldsymbol{X}}^{-1}\mathbf{X}\bar{\boldsymbol{X}}^{-1}\bar{\boldsymbol{V}}  \Bigl]\mathbf{x},
\end{IEEEeqnarray}
i.e.
\begin{IEEEeqnarray}{rCl}
\mathbf{x}^H\mathbf{V}^H\mathbf{X}^{-1}\mathbf{V}\mathbf{x}\geq \mathbf{x}^H\Bigr[\bar{\boldsymbol{V}}^H\bar{\boldsymbol{X}}^{-1}\mathbf{V}+
\mathbf{V}^H\bar{\boldsymbol{X}}^{-1}\bar{\boldsymbol{V}} \nonumber\\ -
\bar{\boldsymbol{V}}^H\bar{\boldsymbol{X}}^{-1}\mathbf{X}\bar{\boldsymbol{X}}^{-1}\bar{\boldsymbol{V}}  \Bigl]\mathbf{x},\
\forall\mathbf{x}\
\end{IEEEeqnarray}
proving (\ref{in1}).

\section*{Appendix B: Proof for inequality (\ref{in2})}
Since function $\ln|\mathbf{X}|$ is concave on $\mathbf{X}\succ  \mathbf{0} $, it is true that
\cite{Tuybook}
\begin{IEEEeqnarray}{rCl}
-\ln|\mathbf{A}|\geq -\ln|\mathbf{B}| - \bigl\la \mathbf{B}^{-1},\mathbf{A}-\mathbf{B}\bigr\ra, \nonumber\\
\forall\ \mathbf{A}\succ \mathbf{0}, \mathbf{B}\succ \mathbf{0}
\end{IEEEeqnarray}
or equivalently
\begin{IEEEeqnarray}{rCl}
\ln|\mathbf{A}^{-1}|\geq \ln|\mathbf{B}^{-1}| - \bigl\la \mathbf{B}^{-1},\mathbf{A}-\mathbf{B} \bigl\ra,\;\nonumber\\
\forall\ \mathbf{A}\succ \mathbf{0}, \mathbf{B}\succ \mathbf{0}. \label{proof28}
\end{IEEEeqnarray}
Then (\ref{in2}) follows by substituting $\mathbf{X}=\mathbf{A}^{-1}$ and $\bar{\mathbf{X}}=\mathbf{B}^{-1}$ into \eqref{proof28}.

\section*{Appendix C: Proof for inequality (\ref{zf8})}
By \cite[Th. 6]{Taetal16} function $\ln (1+x^{-1})$ is convex on $x>0$ so for all $x>0$ and $\bar{x}>0$, it is true that
\cite{Tuybook}
\begin{equation}\label{proof61}
\ln\bigl(1+x^{-1}\bigl) \geq \ln \bigl(1+\bar{x}^{-1}\bigl) + \bigl[(1+\bar{x})^{-1}-\bar{x}^{-1}\bigr](x-\bar{x}).
\end{equation}
Inequality (\ref{zf8}) then follows by substituting $z=x^{-1}$ and $\bar{z}=\bar{x}^{-1}$ into \eqref{proof61}.
\bibliographystyle{IEEEtran}
\balance
\bibliography{NOMAJSAC2017}
\end{document}